\documentclass[12pt]{iopart}
\usepackage{graphicx}

\usepackage{cite}
\usepackage{iopams}

\expandafter\let\csname equation*\endcsname\relax
\expandafter\let\csname endequation*\endcsname\relax
\usepackage{amsmath}
\usepackage{mathtools}
\usepackage [autostyle, english = american]{csquotes}

\begin{document}

\title[Effect of two temperature electrons in a collisional magnetized plasma sheath]{Effect of two temperature electrons in a collisional magnetized plasma sheath}

\author{G. Sharma$^{1}$, S. Adhikari$^{1}$, R. Moulick$^{2}$, S. S. Kausik$^{1,a)}$, and B. K. Saikia$^{1}$}

\address{$^1$Centre of Plasma Physics-Institute for Plasma Research, Nazirakhat, Sonapur-782402, Kamrup, Assam, India}
\address{$^2$Lovely Professional University, Jalandhar Delhi- GT Road, Phagwara, Punjab -144411, India}
%\ead{dutta3dharitree@gmail.com}
\vspace{10pt}
%\begin{indented}
%\item[]April 2019
%\end{indented}
\address{$^{a)}$Email: kausikss@rediffmail.com}

\begin{abstract}
A collisional magnetized plasma consisting of two temperature electrons has been investigated numerically to study the sheath structure and the ion energy flux to the wall. The low-temperature electrons are described by Maxwellian distribution, and the high-temperature electrons are described by truncated Maxwellian distribution. It has been observed that high-temperature electrons play a major role in the sheath potential as well as the ion energy flux to the wall. The presence of collision in the sheath has a significant effect on the properties of the sheath. The study of such a system can help in understanding of plasma surface interaction.
\end{abstract}

%\maketitle

\section{Introduction}

Plasma containing two species of electrons having different temperatures is commonly found in experimental devices\cite{Stangeby, Palop, Yasserian}. For a simple filament discharge in a magnetic multi-dipole device, the primary electrons emitted from the filaments have an approximate temperature of $80~eV$, whereas the equilibrium plasma temperature is less than $10~eV$\cite{Schott}. A high-temperature electron distribution can be achieved by increasing the number of primary electrons, and consequently, a two-temperature electron plasma is produced. In a double plasma device, such a plasma is observed when the plasma potential approaches the cathode potential of the target chamber\cite{Yamazumi}. Besides, two species of electrons are also observed in sputtering magnetron plasma\cite{Sheridan0} and the edge plasma of fusion devices such as tokamak\cite{Stangeby, Eich}. The importance of studying such plasma lies in the plasma surface interaction processes.  Magnetron sputtering is used for sputter etching and thin film deposition. Electrons are trapped by the magnetic field above the cathode, and are responsible for the ion creation. The ions are finally accelerated into the cathode by plasma sheath to cause sputter etching. On the other hand, ion implantation is a technique adopted for modification of surface properties of electrical insulators\cite{Lu, Pelletier, Foroutan}. In this process, the target is kept in plasma and is negatively biased by applying a large negative voltage. Hence, an ion sheath forms around the target. If the plasma contains hot electrons, it changes the ion flux in the sheath, thereby affecting the material properties.

Many researchers\cite{Stangeby, Palop, Yasserian, Yasserian2, Sheridan1, Sheridan2, Gyergyek, Hatami} have studied such systems extensively by considering various models to describe the electrons and ions. Most of the studies are carried out by considering positive ions as fluid and the electrons to be Boltzmann distributed. Schott \cite{Schott}, and Bharuthram \& Shukla\cite{Bharuthram} studied the two temperature electron system for an electrostatic case considering fluid ions. Schott showed that for certain parameters of electron densities (hot electron concentration, $n_{h}<0.276$), there exists more than one solution of plasma edge potential. This leads to the formation of periodic double layer which is similar to the `Bernstein Greene Kruskal' (BGK) equilibrium. On the other hand, Bharuthram and Shukla developed a finite amplitude theory for ion acoustic double layers. Sheridan \textit{et al.}\cite{Sheridan2} studied the plasma containing negative ions and found that multi-layer stratified structure is formed in the sheath for certain values of negative ion density and negative ion temperature. Such a system can be compared with a system containing two temperature electrons because negative ions are considered to be Boltzmann distributed.  Ou \textit{et al.} \cite{Ou}  studied a collisionless two temperature electron plasma system considering the hot electron density as truncated Maxwell Boltzmann and the cold electrons to be Boltzmannian along with fluid ions. They showed that the presence of hot electrons indeed affects the structure of the sheath as well as ion flux to the sheath. Although, they did not find any oscillatory solution of plasma potential. Hatami and Tribeche \cite{Hatami} carried out a study using nonextensive distribution for hot electrons and found that electron temperature ratio has a significant effect on reducing ion velocity at the sheath edge. Yasserian \textit{et al.}\cite{Yasserian} studied plasma sheath with Boltzmann negative ions in the presence of collision and found that the positive ion flux is decreased due to the collision in the sheath.  Sheridan \textit{et al.}\cite{Sheridan0} observed the presence of hot (7.6 eV) and cold (0.5 eV) species of electrons in magnetron plasma. Ikezawa and Nakamura\cite{Ikezawa} found that contamination of a small proportion of the hot electrons enhances the Landau damping of the Bohm-Gross mode, while electron plasma wave propagates. On the other hand, if the temperature difference between the two electron component is sufficiently large, the strength of dispersion is reduced to a significant extent so that the solitary solution is never possible in such plasma\cite{Goswami}.

In contrast to the investigations mentioned above, the presence of ion-neutral collision is supposed to modify the sheath behavior in two temperature electron plasma. In this article, the aim is to study the effect of collision in presence of a magnetic field for a two temperature electron system. The hot species of electrons is described by a truncated Maxwell-Boltzmann and the cold species follows the standard Boltzmann distribution. The paper emphasizes the effect of the ion-neutral collision on sheath formation and its impact on ion energy flux at the wall. The study is carried out over a large range of collision parameters and hot electron concentrations. The paper is organized as follows. Section \ref{model} describes the theoretical formulation of the system, section \ref{criterion} discusses the sheath formation condition along with generation of initial values, section \ref{results} gives the numerical findings and section \ref{concl} contains the concluding remarks.

\section{Theoretical Model}\label{model}
\begin{figure}
		\centering    
		\includegraphics[width=0.7\textwidth]{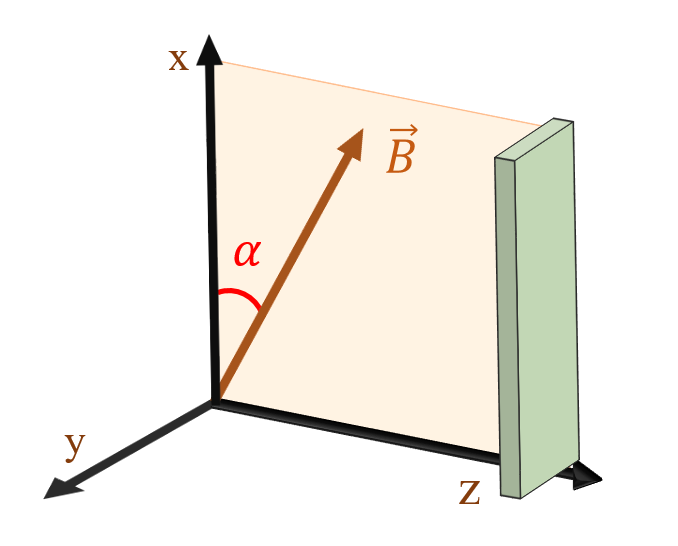}
		\caption[Schematic of the plasma model.]{Schematic of the plasma model.}
		\label{fig:model}
\end{figure}

A collisional magnetized hydrogen plasma is considered with two population of electrons along with positive ions. Figure \ref{fig:model} describes the geometry of the system. The magnetic field lies in the x-z plane making an angle $\alpha$ with the x-axis. The electric field will be along z-axis which is normal to the wall. The fluid ions are singly charged. The electron densities of both hot and cold components vary for different cases. The value of the magnetic field is fixed at $B=0.35T$ and the angle is fixed at $\alpha=10^0$. The plasma is considered to be quasineutral at $z=0$ (bulk plasma).

The cold electron density is described by the Boltzmann distribution

\begin{equation}\label{cold_e_density}
   n_c = n_{c0}~exp\frac{e\phi}{T_c},
\end{equation}

where $\phi$ is the plasma potential and $T_c$ is the cold electron temperature.
For hot electrons, the truncated velocity distribution function is used to incorporate the electron loss at the wall. As the energy of hot electrons is considered to be higher than the wall potential, their absorption at the wall cannot be ignored. The wall is biased at a fixed potential less than the plasma potential so that an ion sheath can form and a cut off speed is defined for the hot electrons\cite{Hutchinson} as $ v_{ch} = \sqrt{2e(\phi(z) - \phi_{wall})/m_e}$, where, $\phi(z)$ is the plasma potential and $\phi_{wall}$ is the biased potential of the wall. Electrons with $v>v_{ch}$ will get absorbed at the wall. Following the concept of cut off speed\cite{Ou, Jelic}, the hot electron density distribution is given by

\begin{equation}\label{hot_e_density}
    n_h =n_{h0}\frac{C_h}{C_{h0}}\exp{\frac{e\phi}{T_h}},
\end{equation}

where, $C_h = \frac{1}{2}(1+erf(\frac{-v_{ch}}{\sqrt{2}v_{th}}))$ is the normalization coefficient, $T_h$ is the hot electron temperature and $v_{th} = \sqrt{T_h/m_e}$ is the electron thermal velocity. Since the hot electrons have easy access to the sheath, there is a possibility that presence of a large number of hot electrons may result in large electron flux at the wall compared to the ion flux $(\Gamma_e>\Gamma_i)$. So formation of sheath may not be possible always in such a case\cite{Hutchinson}. In order to avoid this situation, the hot electrons are considered as minority species.

For the fluid ions, the continuity and momentum equation in 1D is given by 

\begin{equation}\label{cont}
    \frac{d(n_iv_z)}{dz} = S_i,
\end{equation}

\begin{equation}\label{mom}
    v_z\frac{d\textbf{v}}{dz} = -\frac{e}{m_i}\frac{d\phi}{dz}\hat{\textbf{k}} + \frac{e}{m_i}(\textbf{v}\times\textbf{B}) - \frac{S_i}{n_i}\textbf{v} - \frac{1}{m_in_i}\frac{dp_i}{dz}\hat{\textbf{k}} - \nu_i\textbf{v},
\end{equation}

where, $S_i = n_cZ$ is the source term to compensate the loss of ion at the wall, $Z$ is the ionization frequency, $p_i = n_iT_i$ is the ion partial pressure, $n_i$ is the ion density, $T_i$ is the ion temperature and $\nu_i$ is the ion-neutral collision frequency.

The above set of equations are closed by Poisson's equation

\begin{equation}\label{poisson}
    \frac{d\phi}{dz} = -\frac{e}{\epsilon_0}(n_i - n_c - n_h).
\end{equation}

The following normalized variables are used to solve the above equations:

$$\begin{array}{cc}
\xi=\frac{z}{\lambda_{ni}},~~~u=\frac{v_x}{c_s},~~~v=\frac{v_y}{c_s},~~~w=\frac{v_z}{c_s},\\~\\
\lambda_{ni}=\frac{c_s}{Z},~~~N_j=\frac{n_j}{n_{i0}},~~~\eta=-\frac{e\phi}{T_c},~~~\tau_{ic}=\frac{T_i}{T_c},\\~\\
\tau_{hc}=\frac{T_h}{T_c},~~~\gamma_k=\frac{\lambda_{ni}}{c_s}\omega_k,~~~\delta=\frac{n_{h0}}{n_{i0}},~~~K=\frac{\lambda_{ni}}{c_s}\nu_i,~~~a_0=\frac{\lambda_{Di}}{\lambda_{ni}}.
\end{array}$$

Here, $\lambda_{ni}$ is ionization length scale, $c_s=\sqrt{T_c/m_i}$, $\lambda_{Di}=\sqrt{\frac{\epsilon_0T_i}{e^2n_{i0}}}$,  $j=i,c,h$ and $k=x,z$. 

The equations are normalized to ionization length scale.

After normalization, equation (\ref{cold_e_density}) - (\ref{poisson}) take the following forms

\begin{equation}\label{norm_nc}
    N_c = (1-\delta)\exp(-\eta),
\end{equation}

\begin{equation}\label{norm_nh}
    N_h = \frac{C_h}{C_{h0}}\delta\exp(-\eta/\tau_{hc}),
\end{equation}

\begin{equation}\label{norm_flux}
    \frac{d\Gamma_i}{d\xi} = N_c,
\end{equation}

\begin{equation}\label{u}
    \frac{du}{d\xi}=\gamma_z\left(\frac{v}{w}\right) - \left(\frac{N_c}{N_i}\right)\left(\frac{u}{w}\right) - K\left(\frac{u}{w}\right),
\end{equation}

\begin{equation}\label{v}
    \frac{dv}{d\xi}=\gamma_x - \gamma_z\left(\frac{u}{w}\right) - \left(\frac{N_c}{N_i}\right)\left(\frac{v}{w}\right) - K\left(\frac{v}{w}\right),
\end{equation}

\begin{equation}\label{w}
    \frac{dw}{d\xi}=\left(\frac{1}{1-\frac{\tau_{ic}}{w^2}}\right)\left[\frac{1}{w}\frac{d\eta}{d\xi} - \gamma_x\left(\frac{v}{w}\right) - \left(\frac{N_c}{N_i}\right) - K - \left(\frac{\tau_{ic}}{w^2}\right)\left(\frac{N_c}{N_i}\right)\right],
\end{equation}

\begin{equation}\label{norm_poisson}
    \frac{d^2\eta}{d\xi^2}=\left(\frac{\tau_{ic}}{a_0^2}\right)\left(N_i - N_c - N_h\right).
\end{equation}

\section{Sheath criterion}\label{criterion}

In a collisionless unmagnetized plasma, ions must be sonic at the sheath edge in order to form a sheath. This is known as the Bohm criterion for sheath formation. But the presence of magnetic field and collision modifies the situation and sheath may be formed at a subsonic ion speed. The criterion for sheath formation is further modified in the presence of two temperature electrons. A condition for sheath formation based on the present model has been analytically derived. The variation of space charge along $z$ is given by

\begin{equation}\label{sigma}
    \frac{d\sigma}{dz}=\frac{d}{dz}(n_i - n_c - n_h).
\end{equation}

For the formation of the sheath,

\begin{equation}\label{sheath}
    \frac{d\sigma}{dz}>0.
\end{equation}

Now, using equation (\ref{cold_e_density}), (\ref{hot_e_density}), (\ref{cont}) and (\ref{sheath}), the following condition for sheath formation has been obtained

\begin{equation}\label{condition}
\begin{multlined}
    \left[1+\left(\frac{v_y}{v_z}\right)\left(\frac{\omega_x}{\Omega}\right)\right] \\>
    \left(\frac{v_\omega}{v_z}\right)\left[1 - \left(1 - \tau_{ic}\frac{c_s^2}{v_z^2}\right)
    \left(\frac{v_z^2}{c_s^2}\frac{n_c}{n_i} + \frac{v_z^2}{c_s^2}\frac{n_h}{n_i}\tau_{hc} + \frac{v_z^2}{c_s^2}\frac{n_{h0}}{n_i}\tau_{hc}\frac{\exp(e\phi_{wall}/T_h)}{2C_{h0}\sqrt{\pi e(\phi(z)-\phi_{wall})/T_h}}\right)\right].
\end{multlined}
\end{equation}

Here, $v_c=c_s$, $v_h=c_s\sqrt{\tau_{hc}}$, $\Omega=\nu_i+2Z\frac{n_c}{n_i}$, $v_\omega=\frac{eE}{m_i\Omega}$, $E$ is the electric field  and other symbols have their usual meanings. Equation (\ref{condition}) represents a general sheath formation criterion for plasma with two temperature electrons. The idea was adopted from Ref. \cite{Moulick19}. 

The set of equations (\ref{norm_nc}) - (\ref{norm_poisson}) can be solved using a standard RK (Runge-Kutta) scheme. But consistent initial values are needed for each of the unknown variables. The presheath boundary is at $\xi=0$ where the vector quantities have zero values. To estimate the initial values for the situation, the boundary is shifted through an infinitesimally small distance towards the right and a Taylor series expansion of the unknown variables is performed around that point\cite{Moulick17, Moulick19, Forrest, Valentini}. The series is then used in the governing equations to find the coefficients. 

The following series are considered here.

$$\begin{array}{cc}
     N_i = N_{i0}+N_{i1}\xi^2+N_{i2}\xi^4 +~....\\~\\
         \textbf{v}= \textbf{v}_{1}\xi + \textbf{v}_{2}\xi^3 +\textbf{v}_{3}\xi^5 +~....\\~\\
     \Gamma_i = \Gamma_{i1}\xi + \Gamma_{i2}\xi^3 +\Gamma_{i3}\xi^5+~....\\~\\
     \eta = \eta_1\xi^2+ \eta_2\xi^4+ \eta_3\xi^6+~....
\end{array}$$

The first order coefficients of the above series expansions are:

$$\begin{array}{cc}
     \eta_1 = 0, ~~~~\frac{d\eta_1}{d\xi}=0.01,~~~\Gamma_{i1}=N_{c0},\\~\\
     u_1 = \frac{\gamma_x\gamma_zN_{c0}N_{i0}}{\gamma_z^2N_{i0}^2+4N_{i0}^2+4KN_{c0}N_{i0}+K^2N_{i0}^2},\\~\\
     v_1 = \frac{(2N_{c0}+\gamma_x KN_{i0}N_{c0}}{\gamma_z^2N_{i0}^2+4N_{i0}^2+4KN_{c0}N_{i0}+K^2N_{i0}^2},\\~\\
     w_1 = \frac{N_{c0}}{N_{i0}}.
\end{array}$$

From equation (\ref{w}), 
\begin{equation}\label{cond}
    w^2>\tau_{ic}.
\end{equation}

Equation (\ref{cond}) puts a restriction in the value of ion temperature. As the initial values are estimated in the bulk plasma by Taylor series expansion method and the initial value of the velocities are very small, so the ion temperature has to be chosen accordingly so as to meet the requirement. From the initial values for ion velocity, it is seen that they are not fixed and varies for various values of collision parameter, species density, ion mass, and magnetic field.

\section{Results \& Discussions}\label{results}

The major input variables in the present study are the cold and hot electron temperatures, the ion temperature, the hot electron density, and the collision parameter. Throughout the study, the ion temperature is kept constant ($2\times10^{-10}eV$) to meet the condition described by equation (\ref{cond}). The hot electron temperature is also fixed at $50 eV$. Wall potential is maintained at $\phi_{wall}=-20V$. The integration is carried out from presheath edge ($\phi(z)=0$) to the wall ($\phi(z)=\phi_{wall}$).

\subsection{Low $\delta$ and low $K$ regime}\label{low-low}

\begin{figure}
		\centering    
		\begin{minipage}[b]{0.45\textwidth}
		\includegraphics[width=1\textwidth]{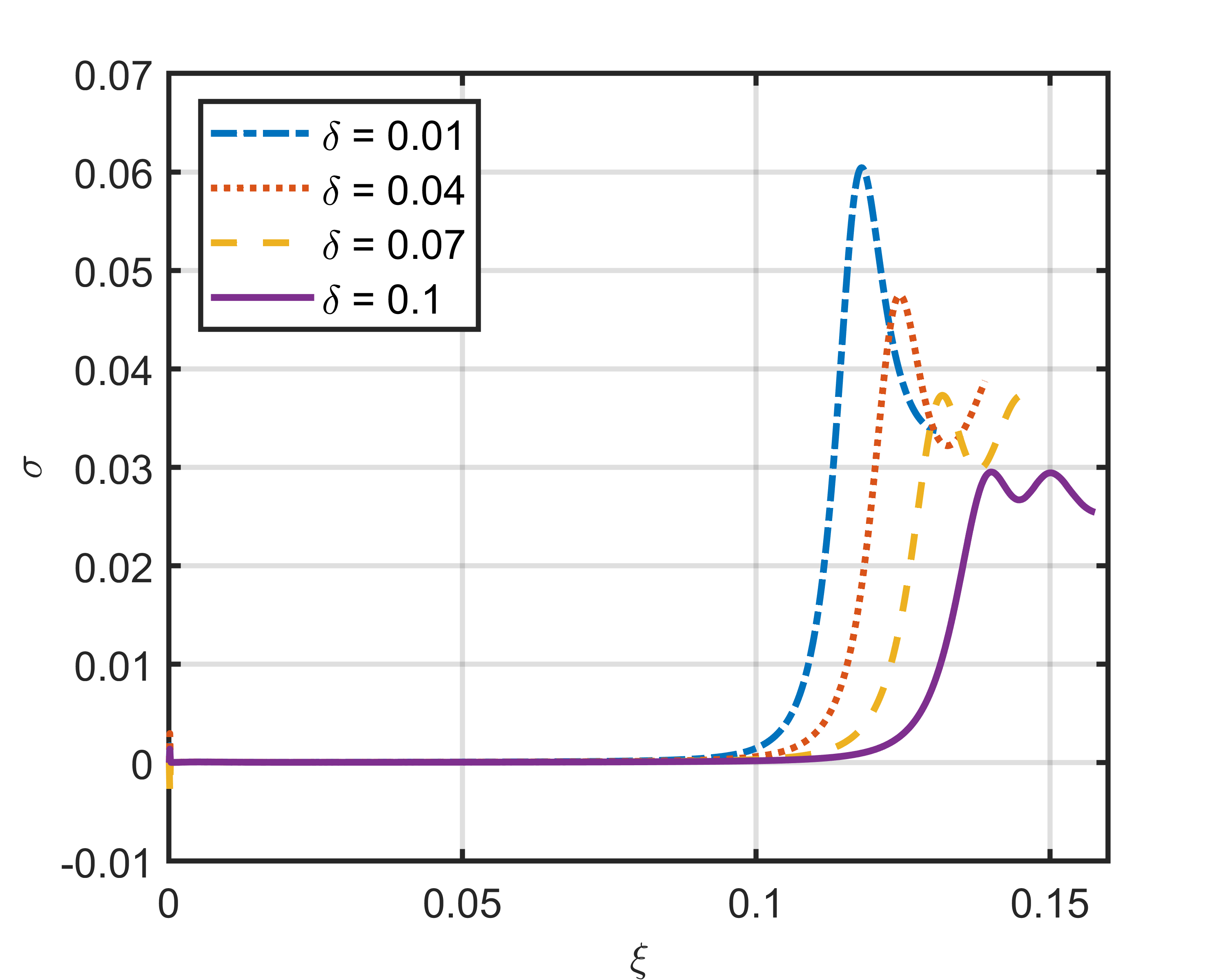}
		\caption[Space charge $(\sigma)$ variation across the sheath with $K=0.1$ for different $\delta$.]{Space charge $(\sigma)$ variation across the sheath with $K=0.1$ for different $\delta$.}
		\label{fig:sigma1}
		\end{minipage}
		\begin{minipage}[b]{0.45\textwidth}
		\includegraphics[width=1\textwidth]{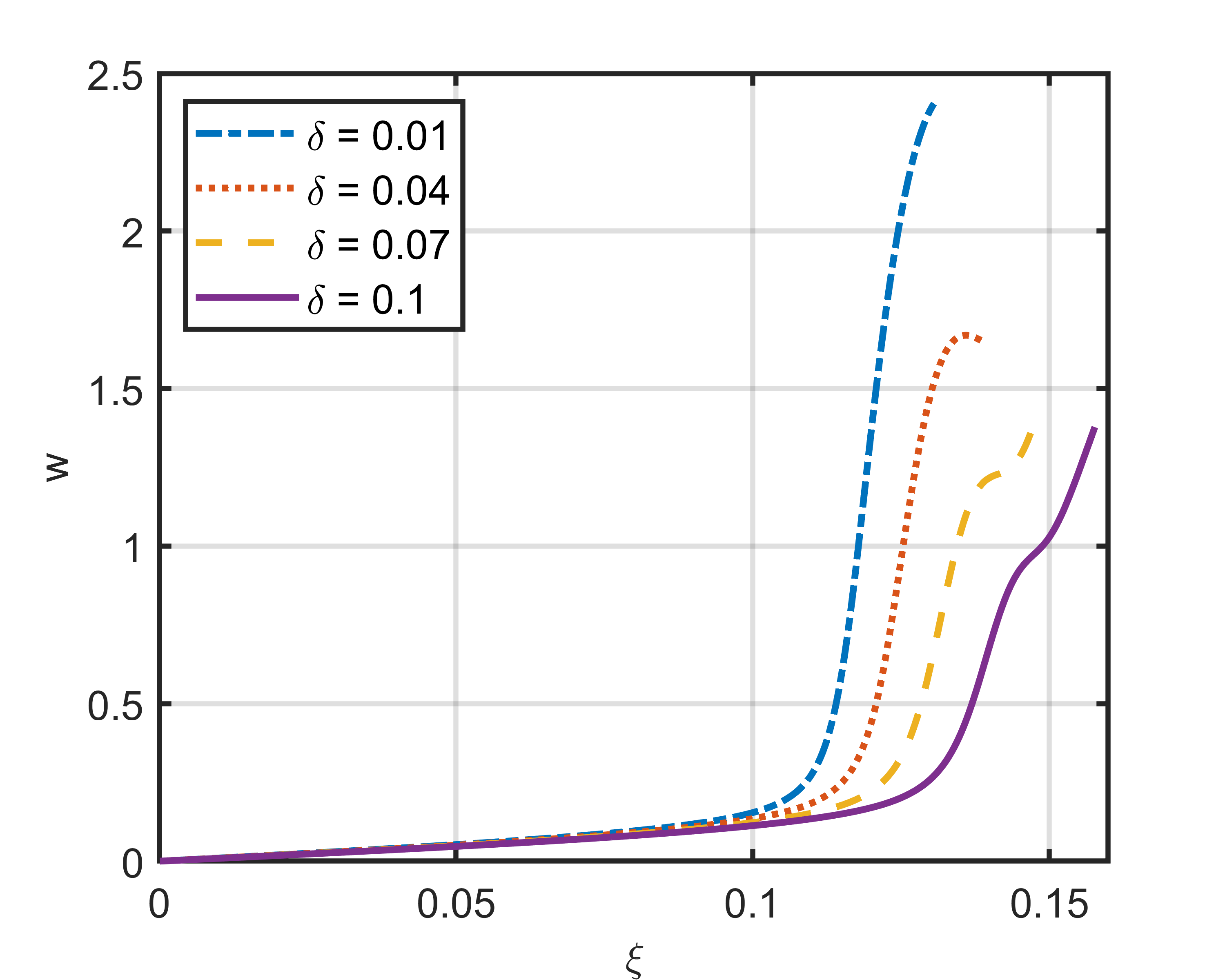}
		\caption[Velocity $(w)$ variation across the sheath with $K=0.1$ for different $\delta$.]{Velocity $(w)$ variation across the sheath with $K=0.1$ for different $\delta$.}
		\label{fig:w1}
		\end{minipage}
\end{figure}

\begin{figure}
		\centering    
		\begin{minipage}[b]{0.45\textwidth}
		\includegraphics[width=1\textwidth]{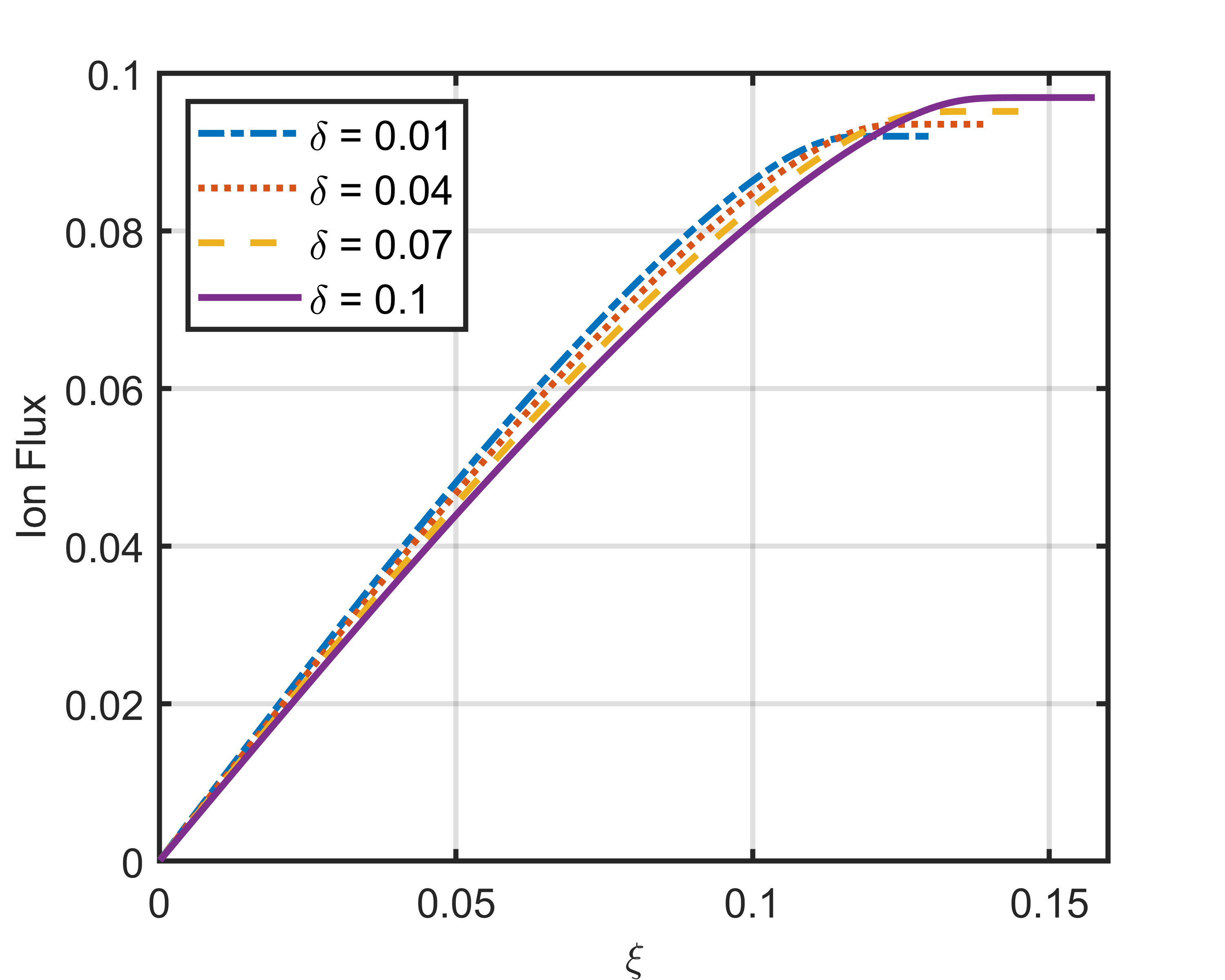}
		\caption[Normalized ion flux $(\Gamma_i)$ variation across the sheath with $K=0.1$ for different $\delta$.]{Normalized ion flux $(\Gamma_i)$ variation across the sheath with $K=0.1$ for different $\delta$.}
		\label{fig:gamma1}
		\end{minipage}
		\begin{minipage}[b]{0.45\textwidth}
		\includegraphics[width=1\textwidth]{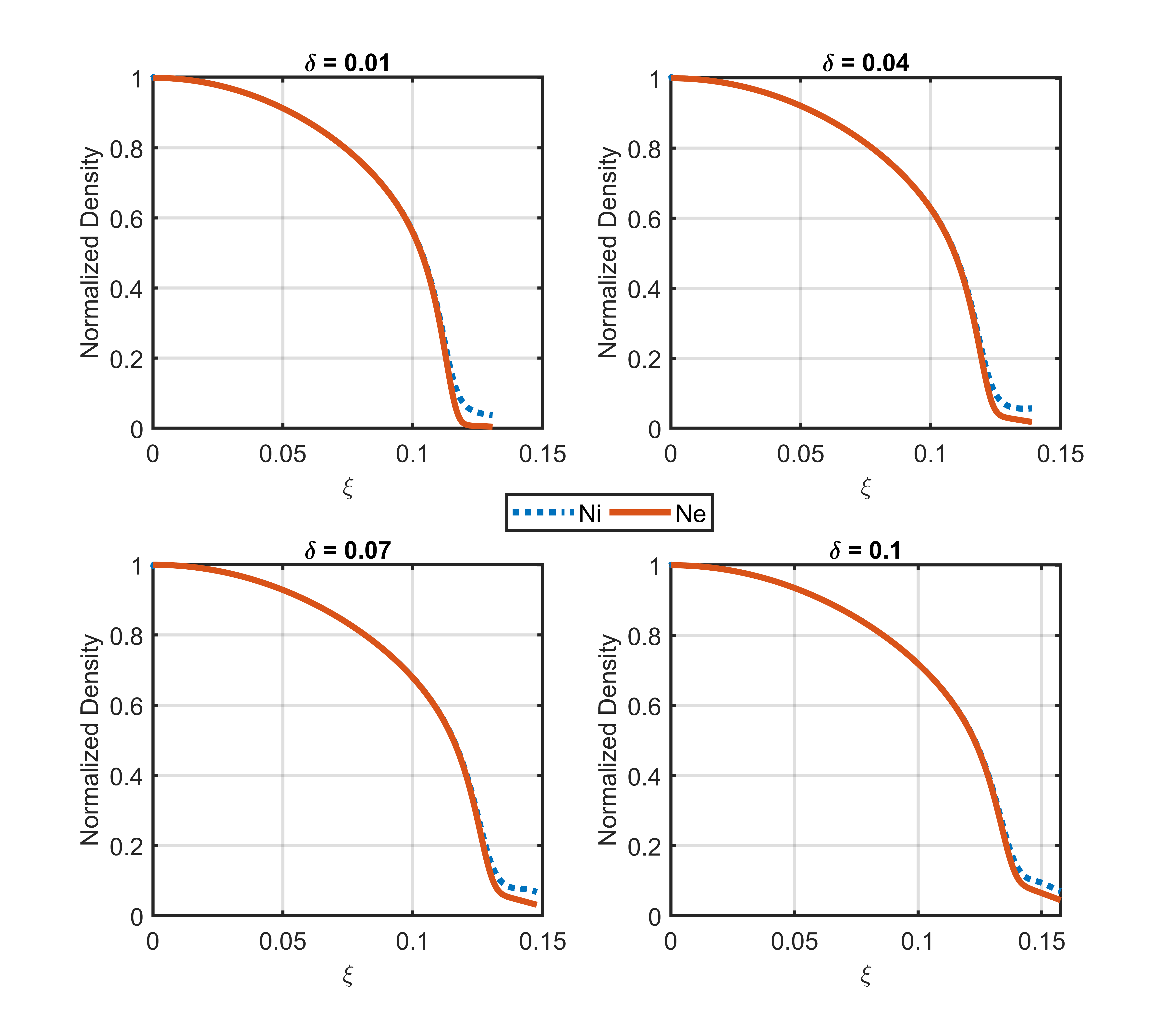}
		\caption[Normalized ion and electron density variation across the sheath with $K=0.1$ for different $\delta$.]{Normalized ion and electron density variation across the sheath with $K=0.1$ for different $\delta$.}
		\label{fig:density1}
		\end{minipage}
\end{figure}

Starting with a lower value of collision parameter, \textit{i.e.}, $K = 0.1$, the hot electron to ion concentration ratio $\delta$ is varied from 0.01 to 0.3. It is observed that there is a noticeable change in the properties of the sheath for the range of $0.01\leq\delta\leq0.1$. In this regime, ionization plays the vital role in the formation of the sheath. The thickness of the sheath gradually increases with an increase of $\delta$. Figure \ref{fig:gamma1} shows that ion flux in the sheath slightly increases with $\delta$. From the profiles of space charge (figure \ref{fig:sigma1}), it is observed that net positive charge in the sheath decreases with an increase in $\delta$. In the ion velocity profiles (figure \ref{fig:w1}), it is observed that ions slightly slow down at a point for higher $\delta$ values, forming a small knee shape. It is the point where ion density attains a local minimum after the formation of the first peak in the space charge profile (figure \ref{fig:sigma1}). The first peak in the space charge is formed at the edge of the sheath where the $z$ component of velocity $(w)$ crosses the characteristic Bohm velocity. The peak gradually shifts towards left of sheath edge as $\delta$ increases. That is $w$ crosses Bohm velocity after the formation of the peak for higher $\delta$ values. The Bohm criterion is modified due to the presence of hot electrons. In this low collisional model, ions attain Bohm speed at the sheath edge for $\delta=0.01$, but as $\delta$ increases, ions enter the sheath with $w<c_s$. Figure \ref{fig:density1} gives the variation in ion and total electron density for various $\delta$.

\begin{figure}
		\centering
		\includegraphics[width=1\textwidth]{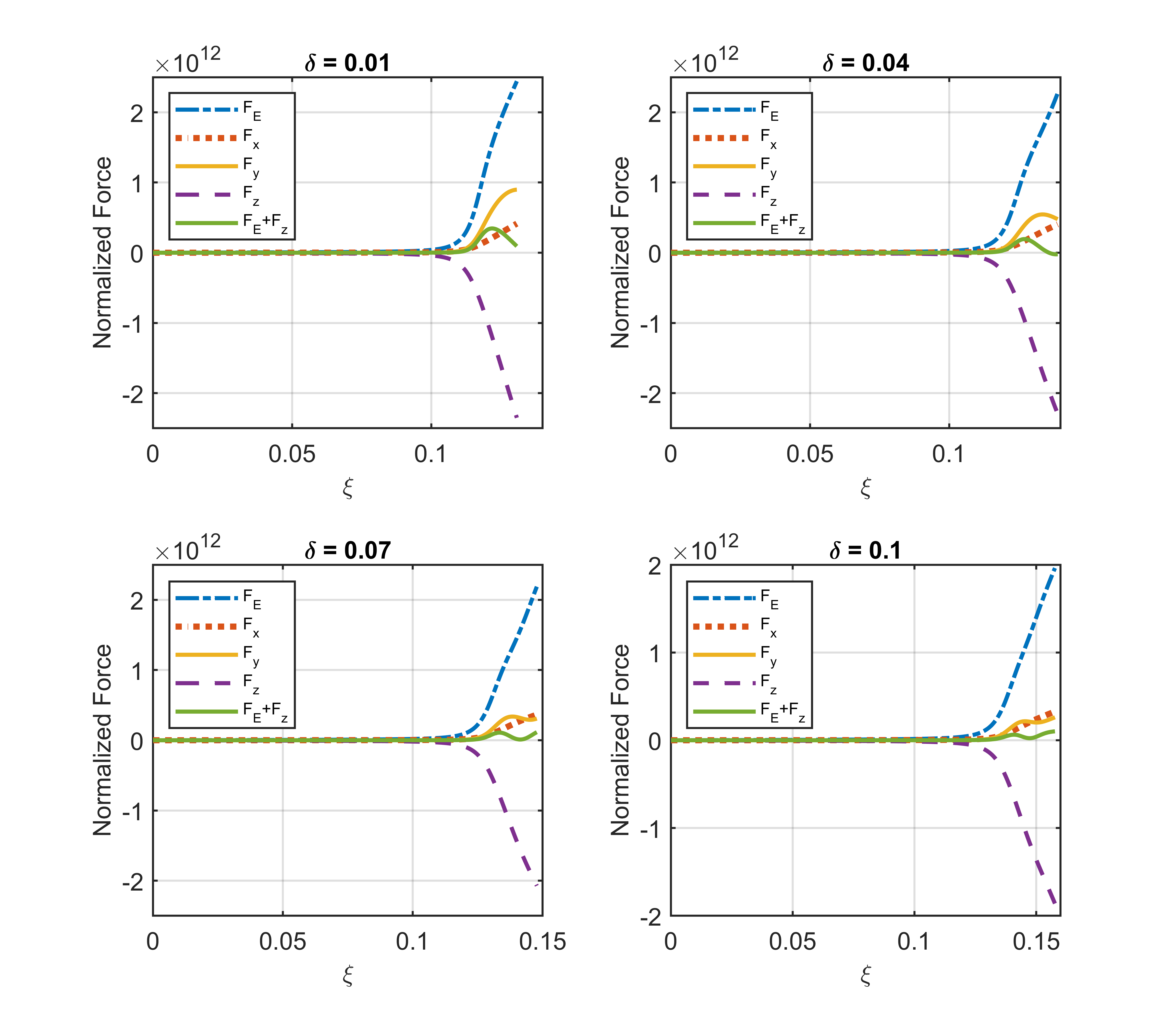}
		\caption[Normalized Lorentz force in the sheath with $K=0.1$ for different $\delta$.]{Normalized Lorentz force in the sheath with $K=0.1$ for different $\delta$.}
		\label{fig:force}
\end{figure}

The above-mentioned observations can be explained as follows: $K<1$ stands for higher ionization frequency over collision frequency. For such a case, ionization becomes the main sheath formation mechanism and it is proportional to cold electron density. If the number of hot electrons in the plasma is significantly large, then the rate of electron absorption at the wall is more likely to be increased. A highly electro-negative wall results in a high electric field, which enables it to attract the ions into the sheath. Hence, the number density of ion in the sheath increases as can be seen in figure \ref{fig:density1}. The flux increases due to low collision and higher ion density in the sheath. As hot electrons have the ability to penetrate through the sheath, with the increase in $\delta$, the net positive charge in the sheath decreases. This is represented by low space charge peaks in figure \ref{fig:sigma1}. The small knee in the profiles of $w$ is due to the Lorentz force. At the sheath edge, the force due to the electric field is balanced by $z$-component of magnetic force as seen in figure \ref{fig:force}. Hence, $w$ decreases at the sheath edge. Inside the sheath, the electric field dominates over the magnetic field for higher values of $\delta$. That is why ions do not significantly slow down as long as $\delta$ is high.

\subsection{Low $\delta$ and high $K$ regime}\label{low-high}

\begin{figure}
		\centering    
		\begin{minipage}[b]{0.45\textwidth}
		\includegraphics[width=1\textwidth]{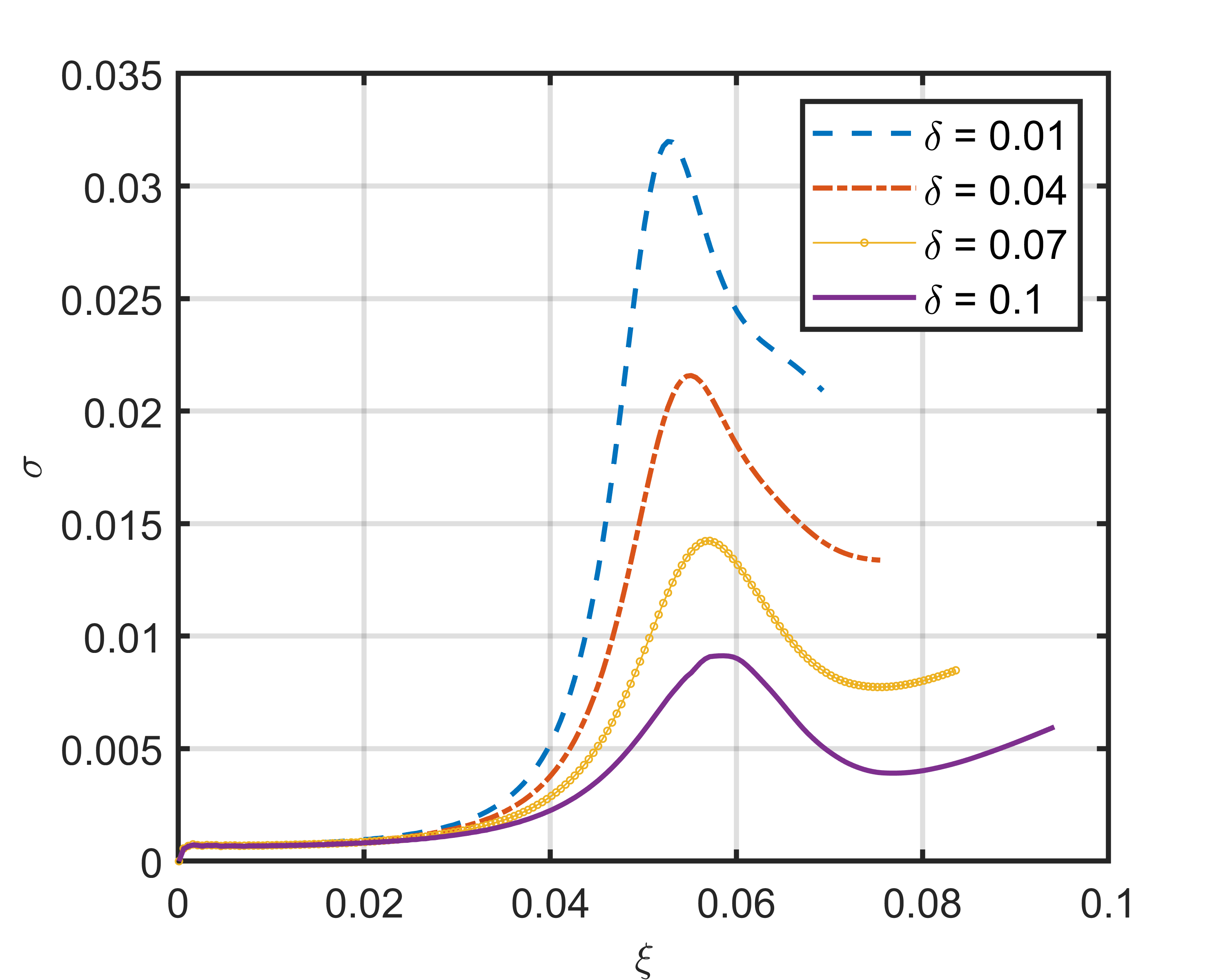}
		\caption[Space charge $(\sigma)$ variation across the sheath with $K=50$ for different $\delta$.]{Space charge $(\sigma)$ variation across the sheath with $K=50$ for different $\delta$.}
		\label{fig:sigma50}
		\end{minipage}
		\begin{minipage}[b]{0.45\textwidth}
		\includegraphics[width=1\textwidth]{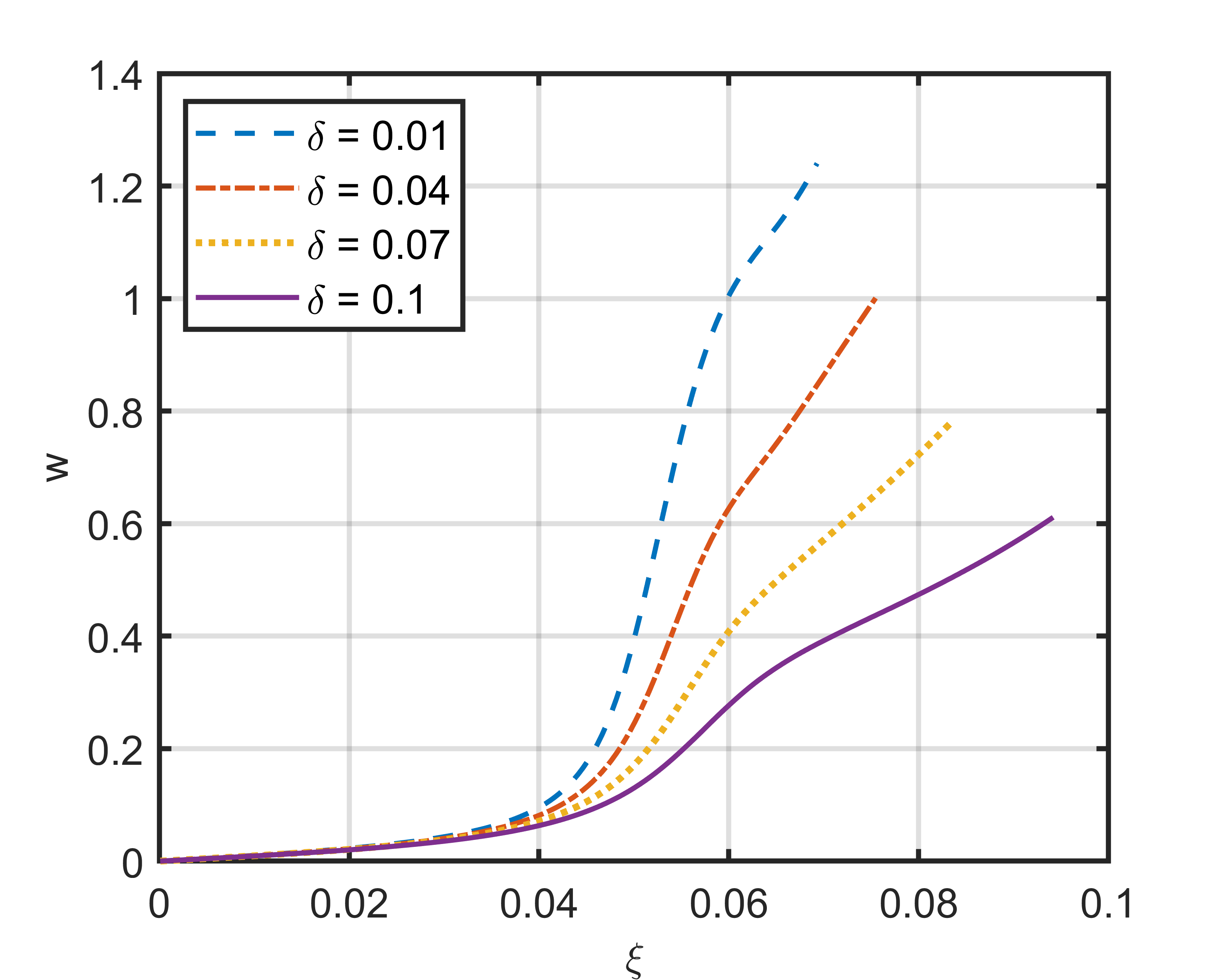}
		\caption[Velocity $(w)$ variation across the sheath with $K=50$ for different $\delta$.]{Velocity $(w)$ variation across the sheath with $K=50$ for different $\delta$.}
		\label{fig:w50}
		\end{minipage}
\end{figure}

\begin{figure}
		\centering
		\begin{minipage}[b]{0.45\textwidth}
		\includegraphics[width=1\textwidth]{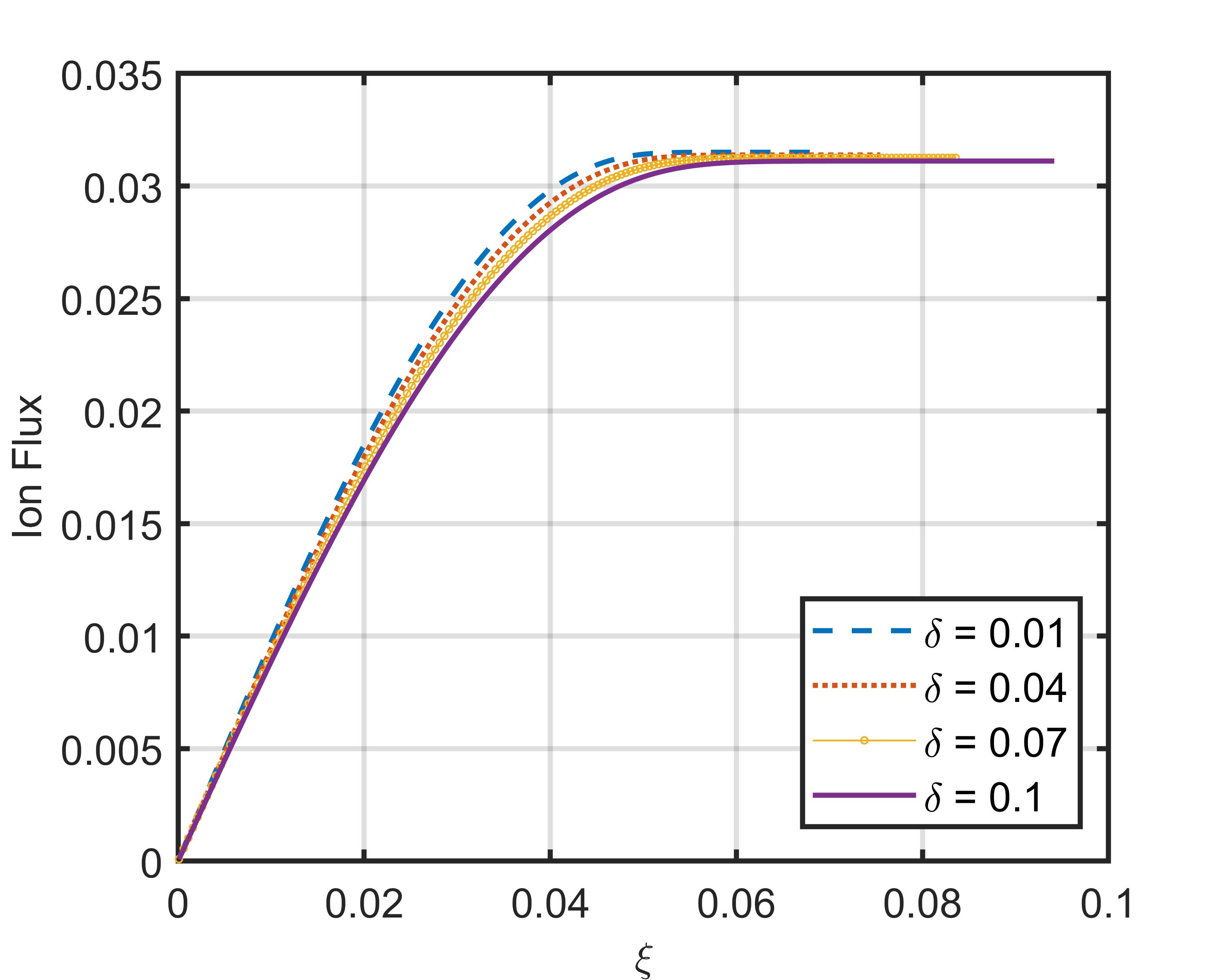}
		\caption[Normalized ion flux $(\Gamma_i)$ variation across the sheath with $K=50$ for different $\delta$.]{Normalized ion flux $(\Gamma_i)$ variation across the sheath with $K=50$ for different $\delta$.}
		\label{fig:gamma50}
		\end{minipage}
		\begin{minipage}[b]{0.45\textwidth}
		\includegraphics[width=1\textwidth]{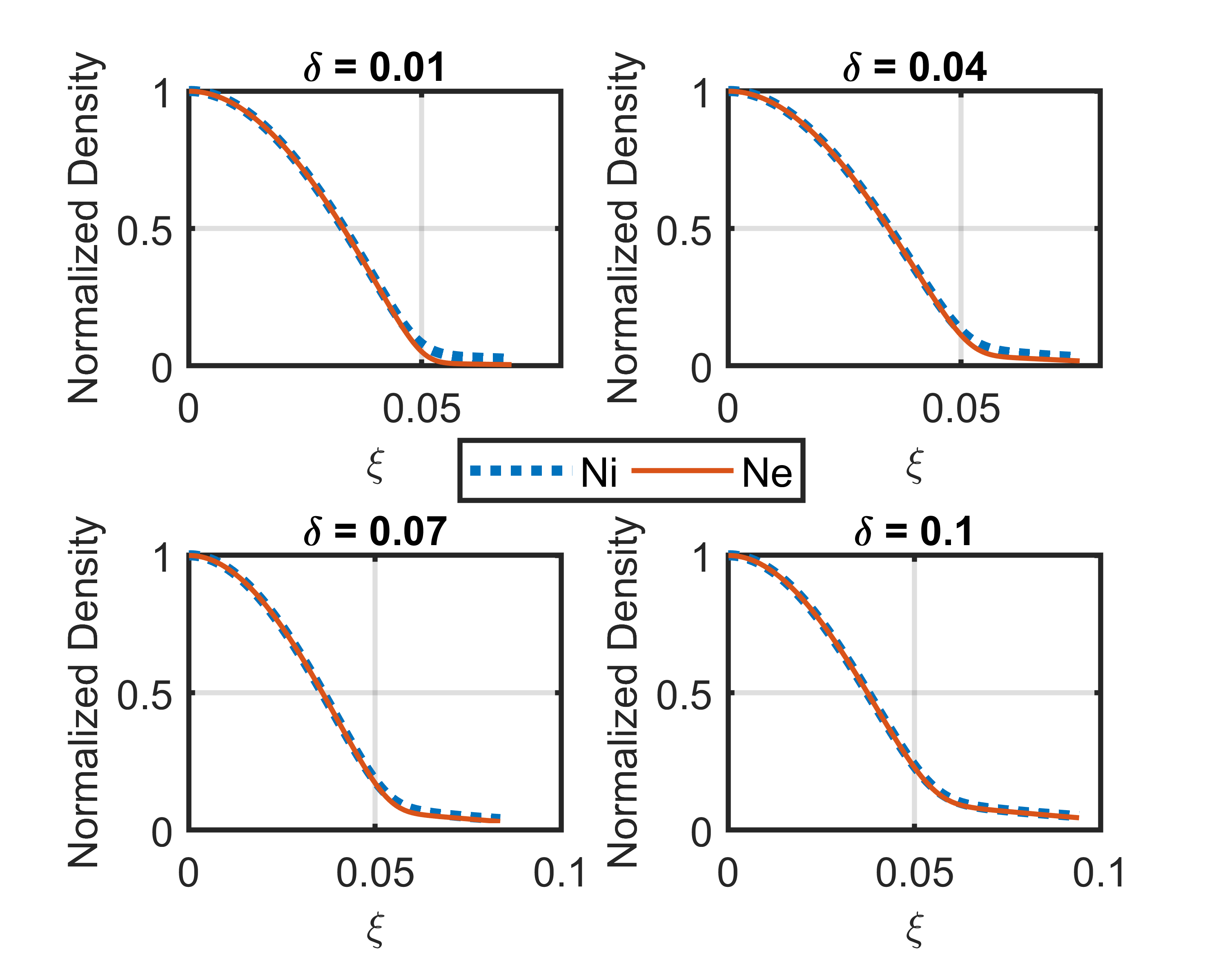}
		\caption[Normalized ion and electron density variation across the sheath with $K=50$ for different $\delta$.]{Normalized ion and electron density variation across the sheath with $K=50$ for different $\delta$.}
		\label{fig:density50}
		\end{minipage}
\end{figure}

When the collision parameter is increased, \textit{i.e.}, $K>1$, which stands for high collision frequency over ionization frequency, the following results are obtained. For $K=50$, figures \ref{fig:sigma50} and \ref{fig:w50} show the space charge and $z$ component of ion velocity. The first thing observed is the disappearance of the bump in the ion velocity. In the space charge profiles, the formation of the secondary peak is not seen. Comparing the plots for ion flux, it is observed that the trend is being changed although insignificant in values, \textit{i.e.}, flux is decreasing for higher values of $\delta$. Another significant observation is regarding the ion velocity $w$ at the sheath edge. So, the presence of collision is affecting the Bohm criterion. Comparing figure \ref{fig:w50} with figure \ref{fig:w1}, it is seen that the velocity gradient decreases with the increase in collision frequency. Ions are moving slowly towards the wall because they are suffering collisions with the neutrals. Hence flux is decreasing as the collision frequency is increased (figure \ref{fig:gamma50}). Figure \ref{fig:density50} displays the ion and electron density variation along $z$ direction for $K=50$.

\subsection{High $\delta$ and low $K$ regime}\label{high-low}

\begin{figure}
		\centering
		\begin{minipage}[b]{0.45\textwidth}
		\includegraphics[width=1\textwidth]{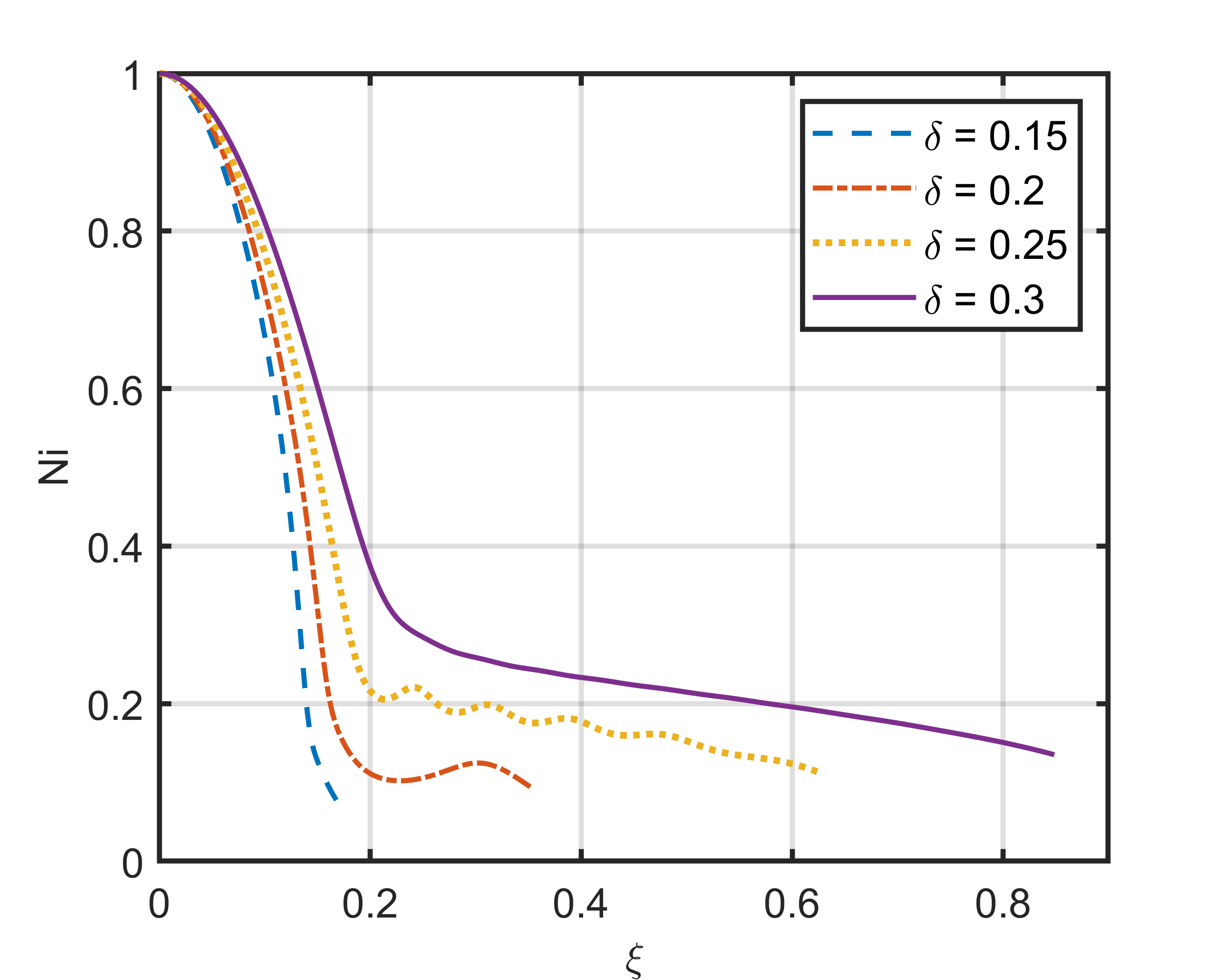}
		\caption[Normalized ion density across the sheath  with $K=1$ for different $\delta$.]{Normalized ion density across the sheath with $K=1$ for different $\delta$.}
		\label{fig:1ni}
		\end{minipage}
		\begin{minipage}[b]{0.45\textwidth}
		\includegraphics[width=1\textwidth]{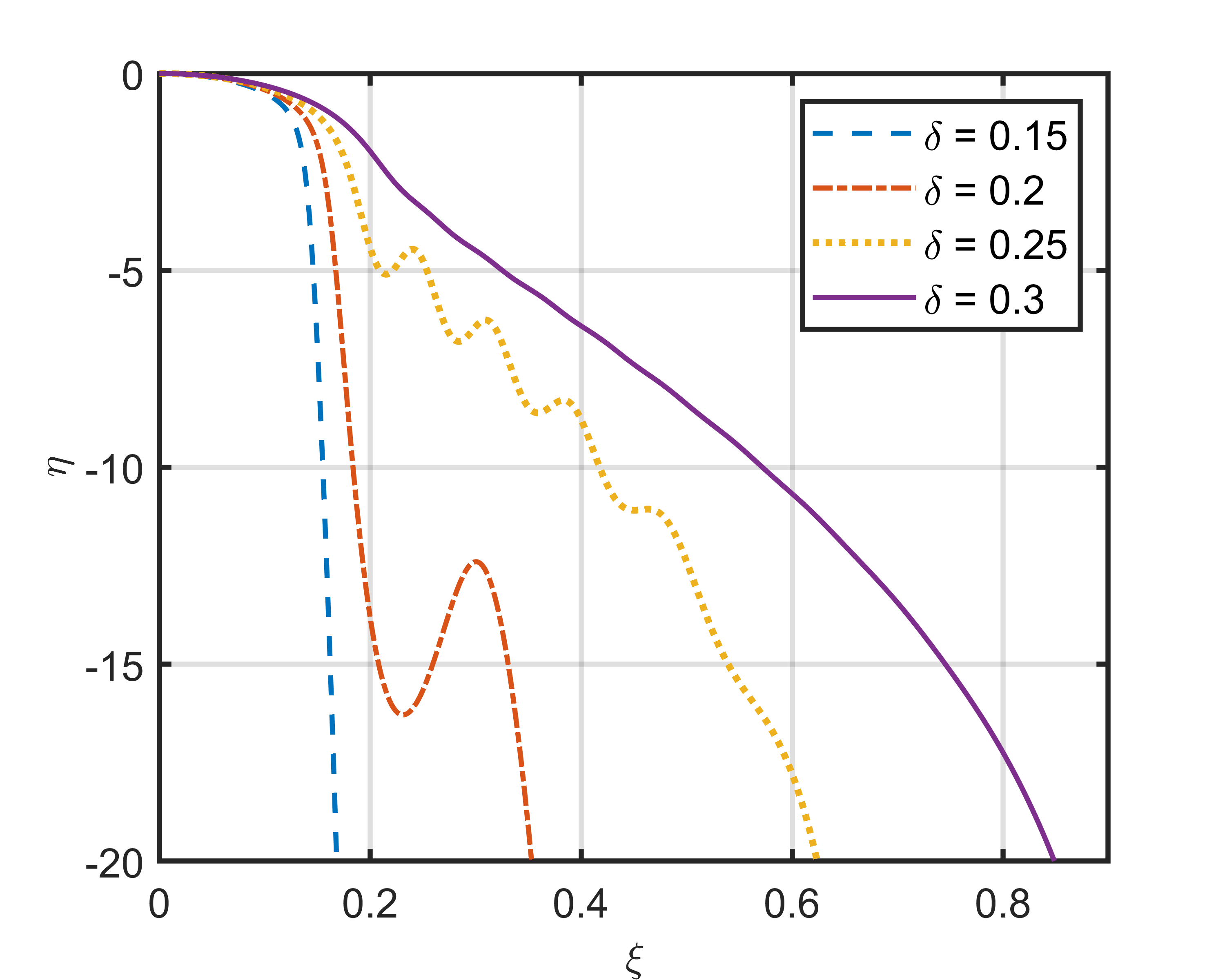}
		\caption[Normalized sheath potential  with $K=1$ for different $\delta$.]{Normalized sheath potential  with $K=1$ for different $\delta$.}
		\label{fig:1pot}
		\end{minipage}
\end{figure}

The value of hot electron concentrations is varied from $\delta = 0.15$ to $\delta=0.3$. The collision parameter is kept between $K=0.1$ and $K=1.0$. It is observed that for a particular range of $\delta=0.2-0.3$, multiple oscillations are observed in the ion density as well as potential profiles. Figures \ref{fig:1ni} and \ref{fig:1pot} represent the ion density and potential variation, respectively for $K=1$. The sheath forms for $\delta=0.15$. For wall potential $\phi_{wall}=-20V$, sheath is not formed for $\delta=0.2-0.3$. The wall potential needs to be more negative to form the sheath after the oscillations in the ion density. 

It has been observed that the occurrence of oscillations in the ion density increases the presheath length scale. Therefore a large spatial length is required to form the sheath. Many authors\cite{Palop, Yasserian, Franklin, Sheridan1, Sheridan2, Yasserian2, Gyergyek2, Braithwaite} have reported such findings in two temperature electron plasma and also for plasma containing Boltzmann negative ions. Such results are termed as multi-layer stratified structure formation and are mentioned as an indication of double layer formation which only occurs for a particular set of parameters\cite{Sheridan1, Sheridan2}. If the fluid model is replaced by kinetic model, for the same parameter regime such results can be obtained\cite{Franklin, Sheridan1, Sheridan2}.

\subsection{High $\delta$ and high $K$ regime}\label{high-high}

\begin{figure}
		\centering
		\begin{minipage}[b]{0.45\textwidth}
		\includegraphics[width=1\textwidth]{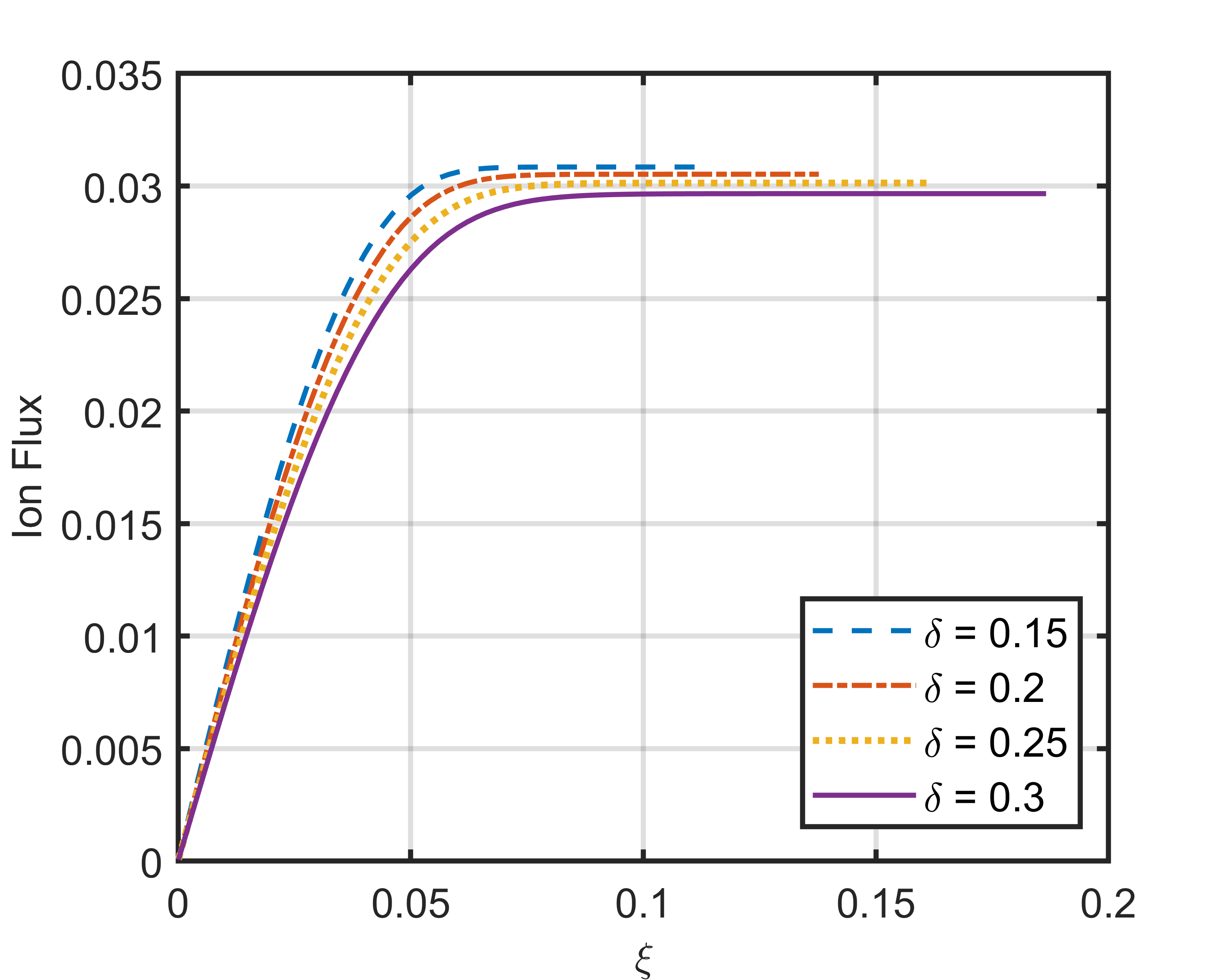}
		\caption[Normalized ion flux across the sheath  with $K=50$ for different $\delta$.]{Normalized ion flux across the sheath with $K=50$ for different $\delta$.}
		\label{fig:flux501}
		\end{minipage}
		\begin{minipage}[b]{0.45\textwidth}
		\includegraphics[width=1\textwidth]{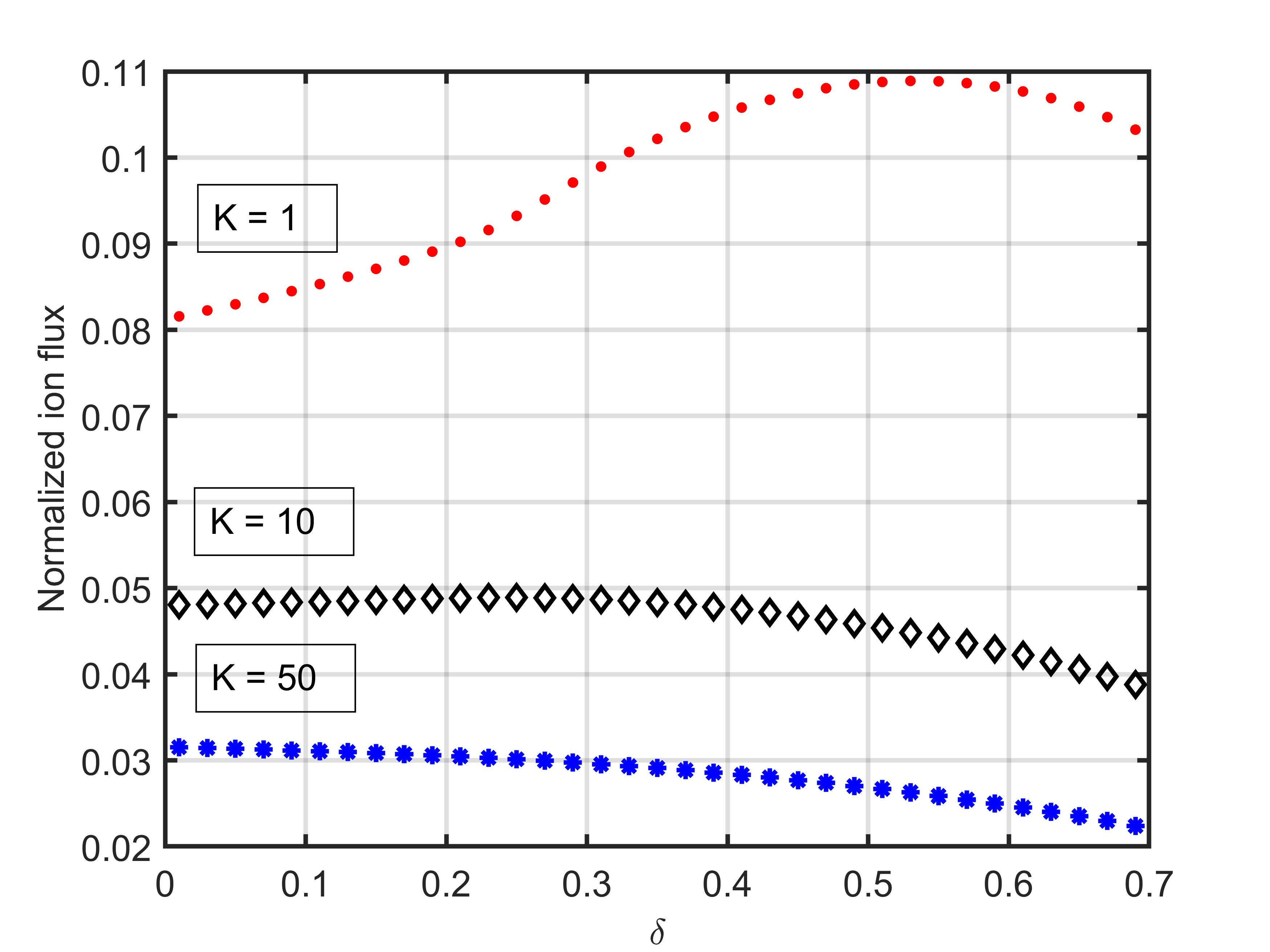}
		\caption[Variation of normalized ion flux at the wall with $\delta$ for different $K$.]{Variation of normalized ion flux at the wall with $\delta$ for different $K$.}
		\label{fig:delta}
		\end{minipage}
\end{figure}

In this case, a higher value $K\geq50$ is chosen and hot electron concentration is varied from $\delta=0.15$ to $\delta=0.3$. No oscillation is found in the ion density profiles. The main observable variation is the change of flux with $\delta$. Previously for low $\delta$ and low $K$ case, it is observed that flux increases with the increase in $\delta$ as long as it remains smaller than 0.1. But for $\delta>0.1$ with high $K$, ion flux decreases with the increase in $\delta$ (figure \ref{fig:flux501}). Also, the net positive charge in the sheath decreases with the increase in $\delta$. Higher $\delta$ stands for more number of hot electrons. Hence, an increase in $\delta$ increases the number of the electron in the sheath, which in turn decreases the net positive charge in the sheath. 

\begin{figure}
		\centering
		\includegraphics[width=0.5\textwidth]{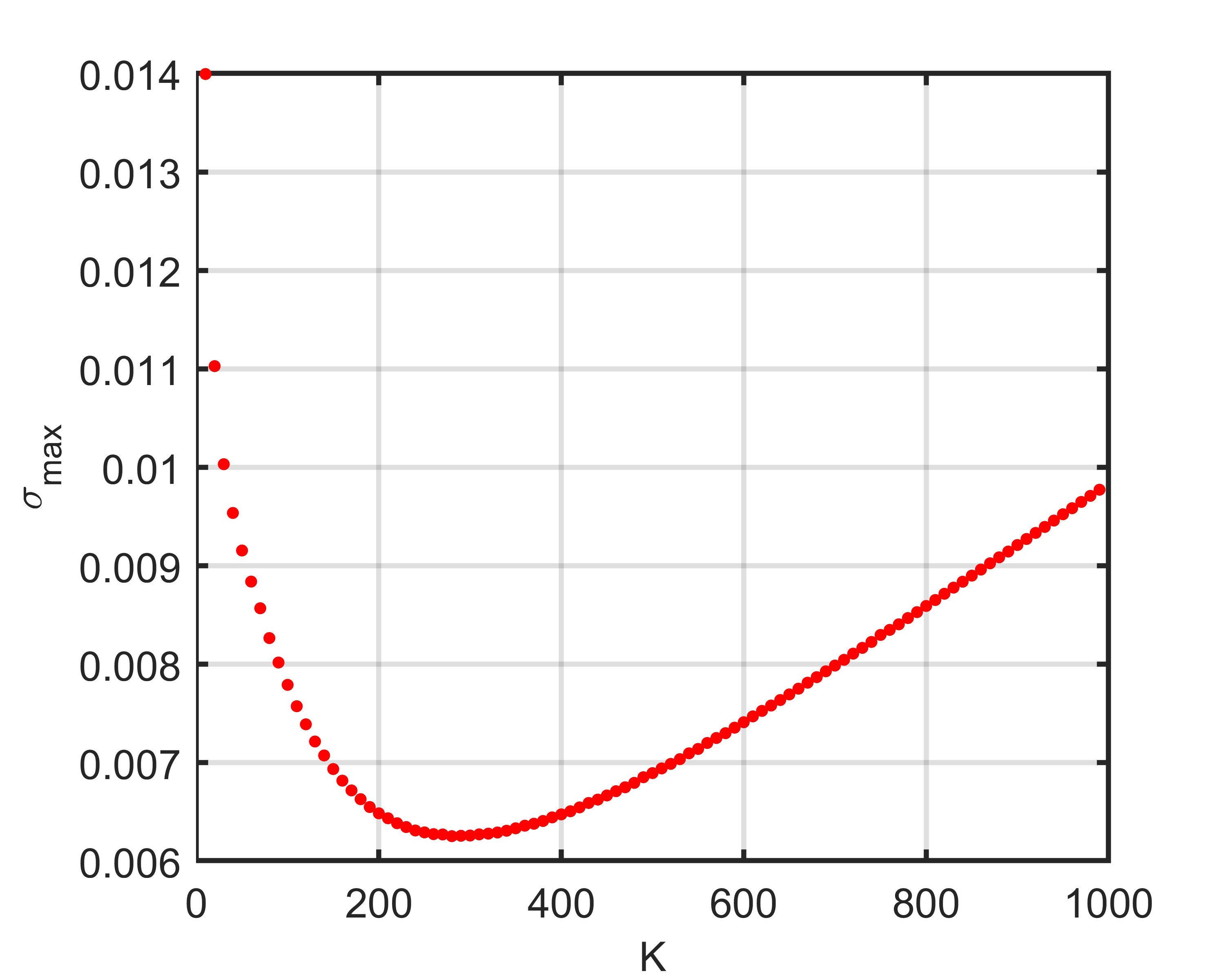}
		\caption[Observed maximum space charge in the sheath for various $K$ values.]{Observed maximum space charge in the sheath for various $K$ values.}
		\label{fig:sigma_max}
\end{figure}

\subsection{Effect of hot electron on ion flux}

To understand the effect of fast electrons on the ion flux in the sheath, $\delta$ is varied from 0.01 to 0.7 for three different values of $K$. Figure \ref{fig:delta} displays the result. It is confirmed that the presence of hot electron is significantly affecting the ion flux. When collision frequency is comparable to ionization frequency, the ion flux is maximum for $\delta = 0.55$. As the collision frequency is increased, ion flux decreases as one moves towards higher values of $\delta$. 

\subsection{Effect of collision on ion velocity and sheath properties}\label{eff_coll}

\begin{figure}
		\centering
		\begin{minipage}[b]{0.445\textwidth}
		\includegraphics[width=1\textwidth]{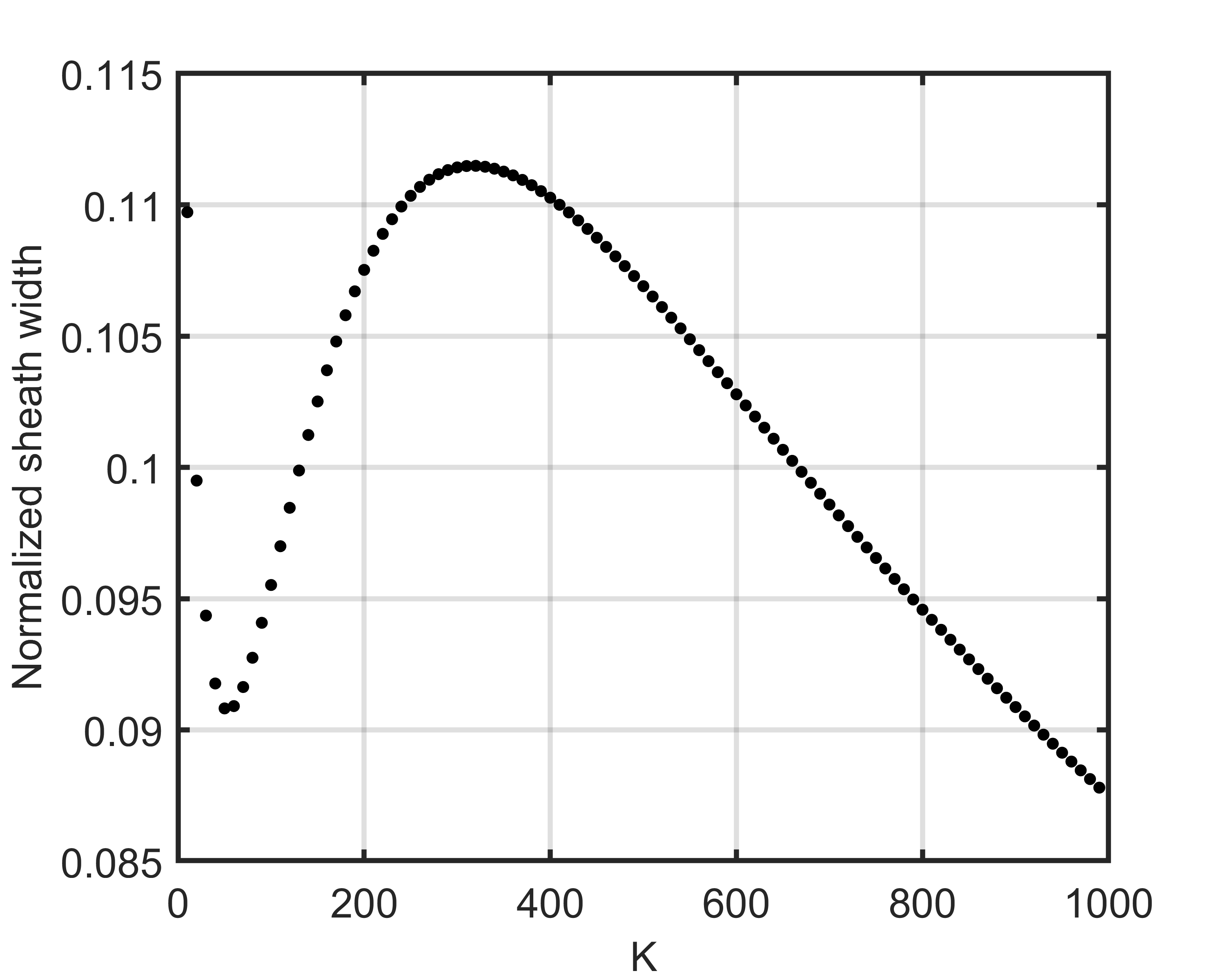}
		\caption[Normalized sheath width  variation for different $K$ values for hydrogen plasma.]{Normalized sheath width  variation for different $K$ values for hydrogen plasma.}
		\label{fig:sheath}
		\end{minipage}
		\begin{minipage}[b]{0.445\textwidth}
		\includegraphics[width=1\textwidth]{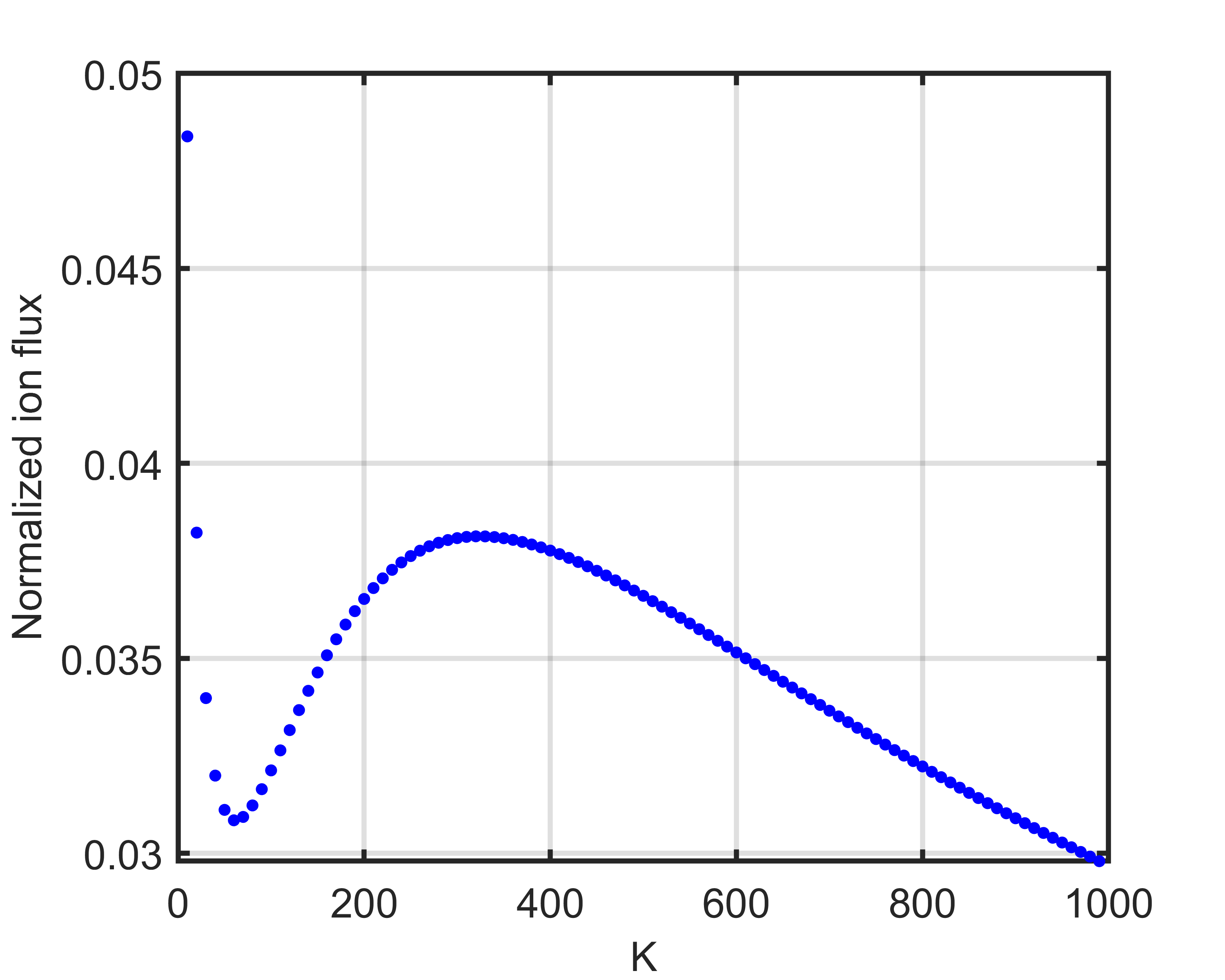}
		\caption[Normalized ion flux in the sheath  for different $K$ values for hydrogen plasma.]{Normalized ion flux in the sheath  for different $K$ values for hydrogen plasma.}
		\label{fig:flux}
		\end{minipage}
\end{figure}

\begin{figure}
		\centering
		\begin{minipage}[b]{0.45\textwidth}
		\includegraphics[width=1\textwidth]{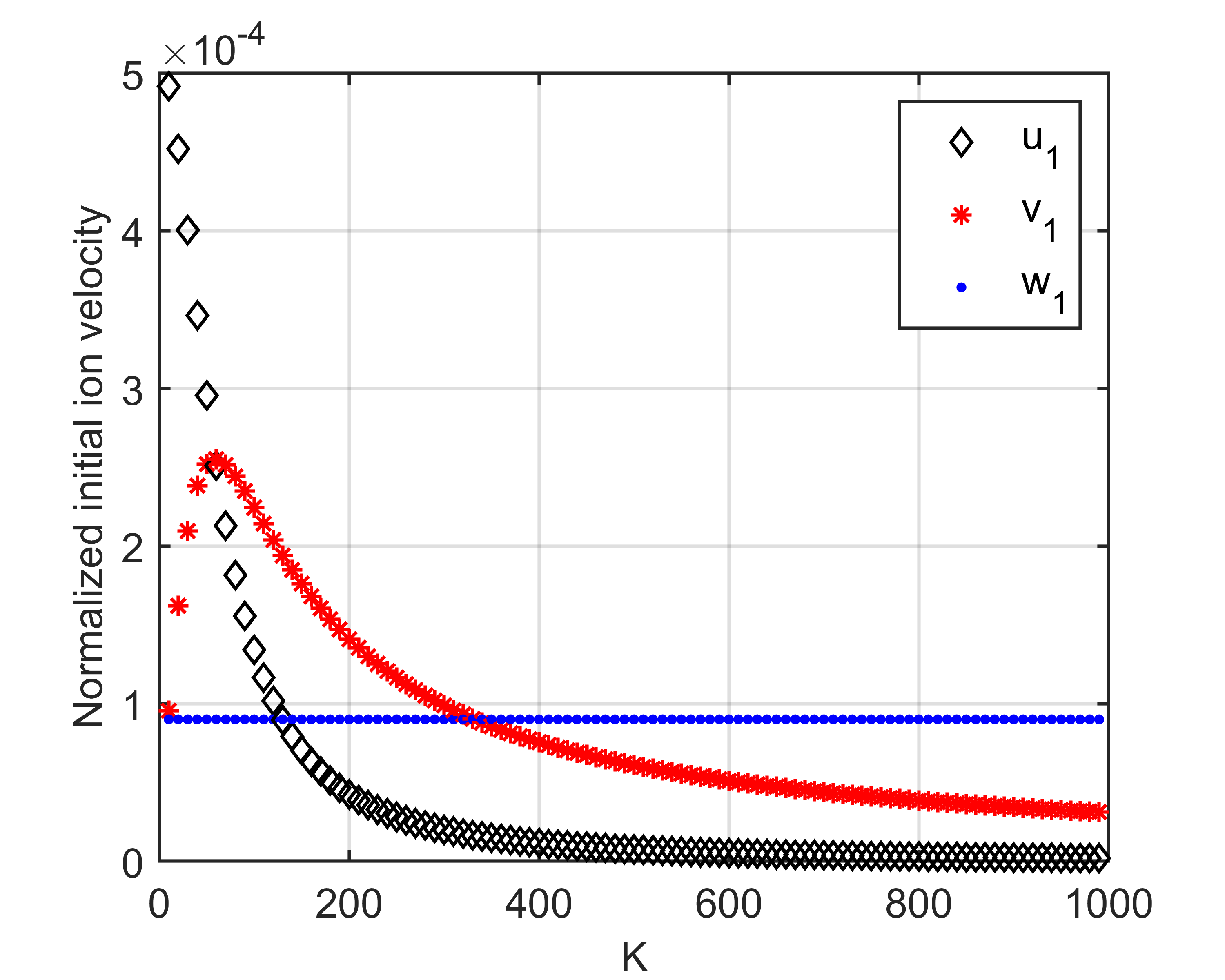}
		\caption[Normalized initial ion velocity variation with $K$.]{Normalized initial ion velocity variation with $K$.}
		\label{fig:initial}
		\end{minipage}
		\begin{minipage}[b]{0.45\textwidth}
		\includegraphics[width=1\textwidth]{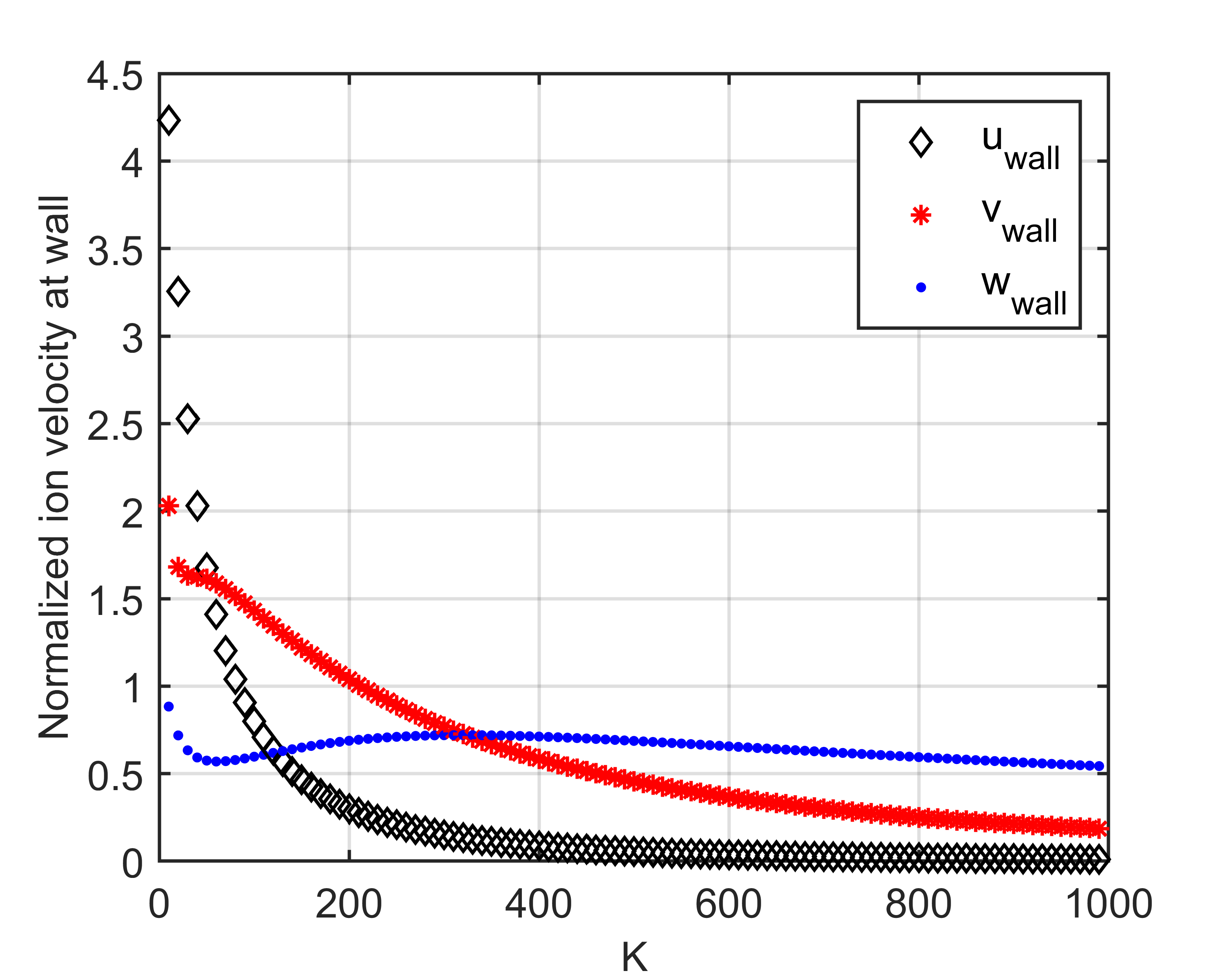}
		\caption[Normalized ion velocity at wall for different $K$ values.]{Normalized ion velocity at wall for different $K$ values.}
		\label{fig:wall}
		\end{minipage}
\end{figure}

After studying the system for fixed values of $K$ and varying $\delta$, the system has been studied for a range of $K$ values keeping $\delta$ constant. To see the effect of the collision parameter on the sheath, the value of $K$ is varied from 10 to 1000 for $\delta=0.1$. It is noticed that the sheath thickness, ion flux, and space charge show variations around two different values of $K$. Figure \ref{fig:sigma_max} shows the variation in maximum space charge in the sheath with $K$, which is found to decrease first, reaches a minimum at $K=300$ and then rises with the increase in $K$. The sheath thickness decreases gradually with the increase in $K$, reaches a minimum at around $K=100$, then rises to a maximum at $K=300$ and then slowly decreases. Figure \ref{fig:sheath} shows the variation of sheath width with $K$.  This pattern is also followed by the ion flux as shown in figure \ref{fig:flux}. Such behavior should be common to all electron-ion plasmas as no departure from these results has been observed when the hot electron concentration is changed. In plotting different components of velocities at the wall (figure \ref{fig:wall}), it is found that in low $K$ regime, $u > v > w$. As K increases $u$ drops to below $v$ and $w$ around $K=100$. Then around $K=300$, $w$ crosses $v$ and maintains the trend for higher values of $K$.


Figure \ref{fig:initial} shows the variation of initial ion velocity with $K$. A similar pattern is also observed for different components of ion velocities at the wall (figure \ref{fig:wall}). In the expressions for initial ion velocities, it is noticed that the $u_1$ and $v_1$ are inversely proportional to $K$. Hence, when  $K$ increases, $u_1$ and $v_1$ decreases. But $w_1$ is independent of $K$. So, at two particular values of $K=100$ and $K=300$, $u_1$ and $v_1$ becomes less than $w_1$. Now the reason behind this particular behavior around $K=100$ and $K=300$ is that the x-component of ion gyrofrequency $\omega_x$ becomes approximately equal to collision frequency, $\nu_i$ for $K=300$ and $z$-component of gyrofrequency $\omega_z$ becomes approximately equal to collision frequency for $K=100$. The x-component of force $F(x)$ depends on $\omega_z$ and $\nu_i$. Around $K=100$, these two main components nullify their effects because of which there is a steep fall of $u$ around $K = 100$. The y-component of force $F(y)$ depends on $\omega_x$, $\omega_z$ and $\nu_i$. Around $K = 300$, the term containing $\omega_x$ cancels the effect of collision term, because of which force decreases and $v$ also decreases. But for z-component of force $F(z)$, collision term and gyro frequency term share the same sign, hence no cancellation occurs and $w$ increases. But a further increase in $K$ decreases the ion velocity. As $w$ attains local maximum around $K = 300$, hence flux also maximize locally. when the collision parameter $K$ has a moderate value, the $y$-component of velocity $v$ has the highest sharing among the three. This is because of the $\textbf{E}\times\textbf{B}$ drift force. But for high $K$ values, collision force dominates and hence $u$ and $v$ decreases. Figure \ref{fig:sigma_max} shows the variation of maximum space charge observed in the sheath for different $K$. As mentioned earlier, the trend is changed at around $K=300$. For $K<300$, it is gradually decreasing and for $K>300$, it increases with an increase in $K$.

\subsection{Effect of ion mass on the sheath}\label{mass}

\begin{figure}
		\centering
		\begin{minipage}[b]{0.445\textwidth}
		\includegraphics[width=1\textwidth]{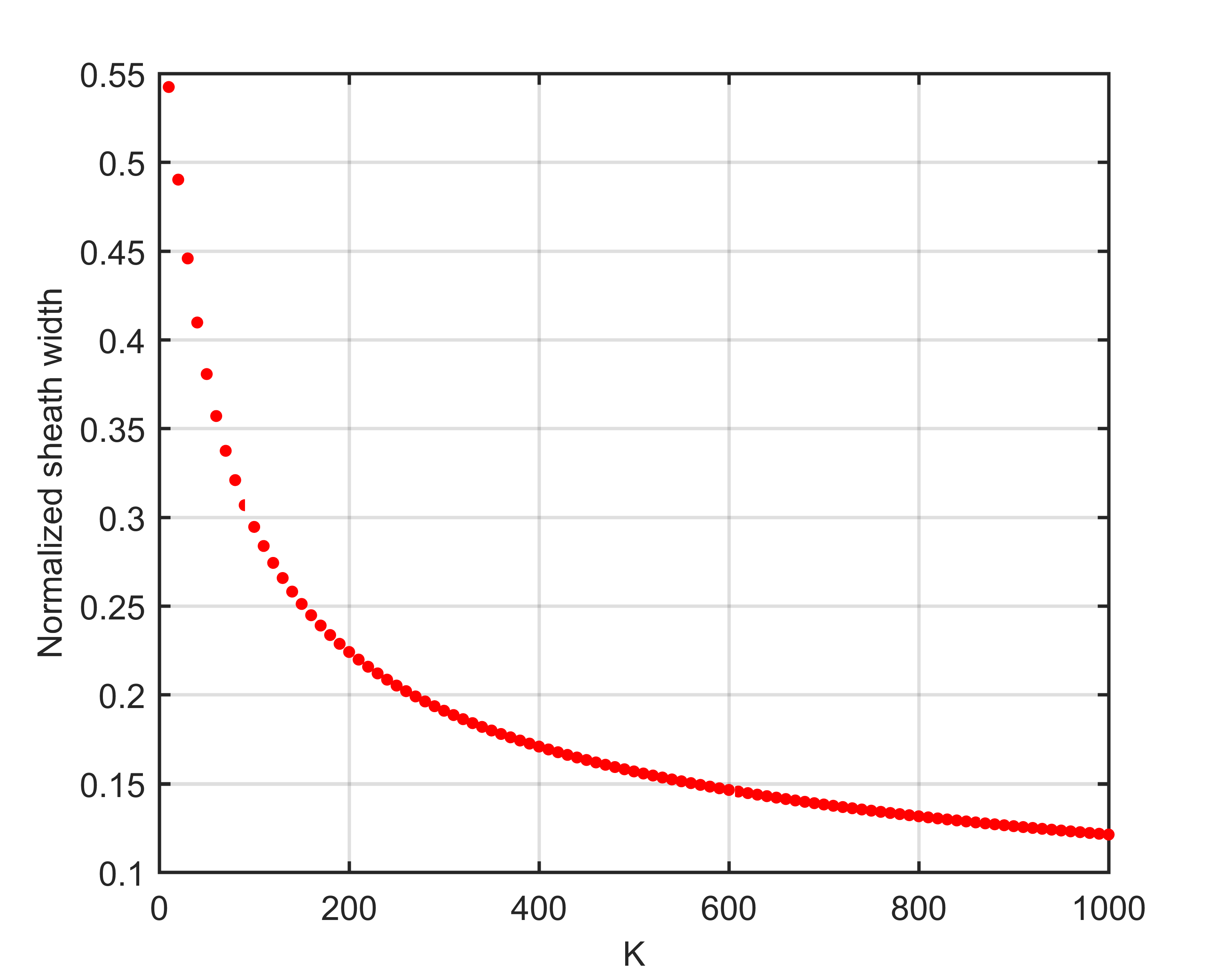}
		\caption[Normalized sheath width  variation for different $K$ values for argon plasma.]{Normalized sheath width  variation for different $K$ values for argon plasma.}
		\label{fig:sheath2}
		\end{minipage}
		\begin{minipage}[b]{0.44\textwidth}
		\includegraphics[width=1\textwidth]{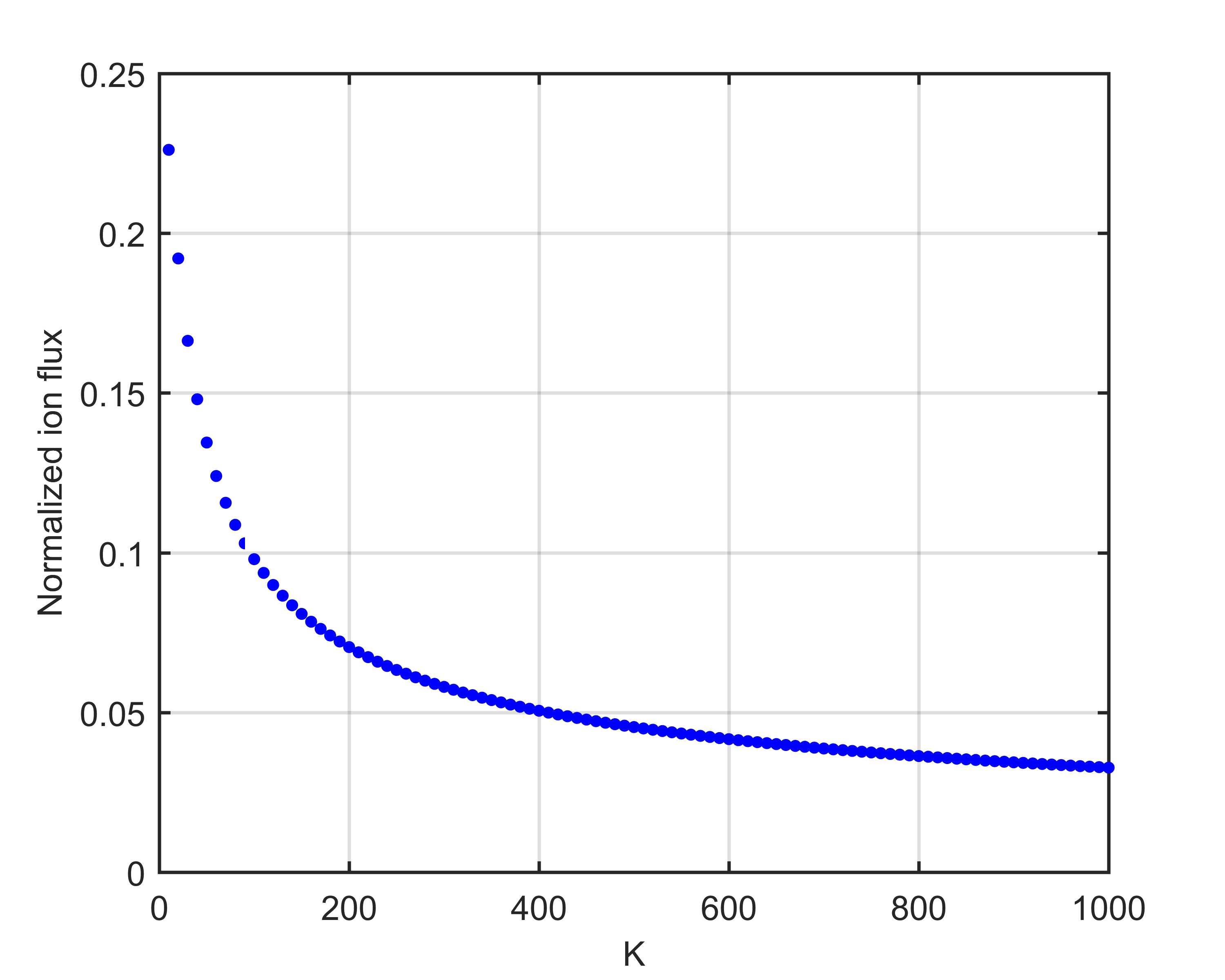}
		\caption[Normalized ion flux in the sheath  for different $K$ values for argon plasma.]{Normalized ion flux in the sheath  for different $K$ values for argon plasma.}
		\label{fig:flux2}
		\end{minipage}
\end{figure}

To confirm the results obtained in the previous section, the mass of the ion is increased to 40 AMU which corresponds to an argon plasma. For this case, the thickness of the sheath and ion flux on the wall is investigated for a range of $K$ values. Figures \ref{fig:sheath2} and \ref{fig:flux2} display the results. It is observed that unlike the previous case (figures \ref{fig:sheath} and \ref{fig:flux}), the sheath thickness and the ion flux decreases with an increase in $K$. For argon plasma, the ion gyrofrequency is greater than the range of collision frequency that is being considered here. Therefore, no cancellation of forces occurs as seen in the previous case and hence the ion flux and sheath width smoothly decrease with $K$.

\section{Conclusions}\label{concl}

Using a single fluid approach, sheath formation in collisional magnetized plasma is studied in the presence of two temperature electrons. It is found that the presence of hot electron changes the sheath properties. For a particular set of $K$ and $\delta$, the potential becomes multivalued and ion density shows oscillatory behavior where no sheath is formed. In this regime, relatively higher negative wall potential is required for sheath formation. Apart from that, the ion-neutral collision frequency plays an important role in the formation of the sheath. When collision frequency increases linearly, the sheath width gradually decreases. For a particular choice of input parameters, when $\nu_i$ and $\omega$ becomes equal, the ion flux and sheath width reach maximum irrespective of hot electron concentration. The present study shows that the properties of the sheath are non-monotonic functions of ion-neutral collision. If the processing plasma is collisional, the study shows that to get maximum ion flux, the magnetic field has to be chosen in such a way that the ion gyrofrequency becomes equal to the collision frequency. On the other hand, it would always be better to minimize the hot electron concentration for a collisional plasma in order to achieve maximum ion flux. Results of the study may be experimentally realized in a magnetic multi-dipole device, where two temperature electrons can be produced.

\section*{Acknowledgements}

The study has been carried out with the financial support from the Department of Atomic Energy, Government of India.

%*****************************************************************************************	

\end{document}